\documentclass[aps,preprint,amsmath,amssymb,showpacs]{revtex4-1}
\usepackage{amsmath}
\usepackage{amssymb}
\usepackage{graphicx}
\usepackage{epstopdf}
\usepackage{float}
\usepackage{subfigure}
\usepackage{color}
\usepackage{hyperref}
\usepackage{comment}
\usepackage[utf8x]{inputenc}

\begin{document}
\title{Effects of probe size, shape, and probe-surface interactions in imaging multilayer island growth, coalescence, and formation of a smooth film}
\author{Ismael S. S. Carrasco}
\email{ismael.carrasco@ufv.br}
\affiliation{Departamento de F\'isica, Universidade Federal de Vi\c cosa, 36570-900, Vi\c cosa, MG, Brazil}
\author{Tiago J. Oliveira}
\email{tiago@ufv.br}
\affiliation{Departamento de F\'isica, Universidade Federal de Vi\c cosa, 36570-900, Vi\c cosa, MG, Brazil}
\author{F\'abio D. A. {Aar\~ao Reis}}
\email{reis@id.uff.br}
\affiliation{Instituto de F\'isica, Universidade Federal Fluminense, Avenida Litor\^anea s/n, 24210-340 Niter\'oi, Rio de Janeiro, Brazil}
\date{\today}

\begin{abstract}

Recent studies of deposition onto weakly interacting substrates have revealed multilayer island growth, followed by smooth-film formation upon island coalescence. However, for $\text{C}_{60}$ films, the roughness peak measured by atomic force microscopy (AFM) was significantly lower than that obtained by X-ray reflectometry (XRR), suggesting tip–sample convolution effects. We investigate this problem using kinetic Monte Carlo (KMC) simulations of such films and models of probe–surface interactions for AFM imaging modes. To model the non-contact mode, surface profiles are transformed by averaging heights over regions matching the probe-tip width. A roughness peak remains if the probe width $\ell$ is smaller than the average island distance, but its height decreases with $\ell$ and its position is slightly displaced. In the contact-mode model with cylindrical probes and hard-core interactions, the peak also decreases and shifts significantly to smaller thicknesses, indicating apparently faster island coalescence. Thus, SPM images under these conditions may underestimate film-thickness nonuniformity in deposits with coalescing islands. Instead, models of sharp pyramidal or conical probes with hard-core interactions produce deviations in the roughness peaks only for large $\ell$, although such sharp tips are difficult to realize experimentally. A geometric model of cuboid island growth and coalescence quantitatively confirms the effects of contact modes and probe shapes observed in the simulations. The results suggest that post-growth processes, such as island coarsening and dewetting, may have occurred in the recently studied $\text{C}_{60}$ film between the XRR and AFM measurements.

\end{abstract}

\maketitle

\section{Introduction}
\label{secIntro}

The fabrication of heterostructures for electronic, photovoltaic, photocatalytic, and battery technologies requires the growth of thin solid films of different materials for performing a variety of functions needed in these devices \citep{ohring,Grundmann2021,qiaoACSANM2023,ranaFuel2024,artuk2024}.
As a film is deposited on the surface of another material, i.e., on a foreign substrate, the formation of multilayer islands is expected if the interactions between the deposited molecules are stronger than the film-substrate interactions \citep{ohring,pimpinelli,volmerweber}.
The increase of the island heights concomitantly with their lateral widening implies an increase of the surface roughness (defined as the rms fluctuation of the local thickness).
Although many film-substrate combinations lead to this growth mode, it is usually undesired because a high density of grain boundaries can be formed when the islands coalesce and high (tensile or compressive) stress can develop \citep{kochPRL2005,yuActaMater2014,abadias2015,huttner2019}.
Moreover, when the multilayer islands coalesce and form a continuous film, different morphological evolutions may be observed.
In the deposition of several materials from vapor or with electrochemical methods, a monotonic increase of the roughness is observed, possibly with the formation of a plateau during that coalescence; this is the case, e.g., in films of copper \citep{guoElecComm2010}, tantalum \citep{yangJAP2012}, cobalt phthalocyanine \citep{gedda2014}, and $\text{WS}_2$ \citep{pradhan2020}.
However, some studies with inorganic and organic materials show a decrease of the roughness as the islands coalesce and give rise to a reasonably smooth film surface, which is the so called rough-to-smooth growth mode; this is the case of films of $\text{Fe}_3\text{Si}$ \citep{kaganer2009}, CdTe \citep{toApplSS2021}, $\text{C}_{60}$ \citep{lapkinPRL2026}, and 1,4,5,8,9,11-hexaazatriphenylene-hexacarbonitrile (HATCN) \citep{lapkinPRL2026} deposited on a variety of substrates.
Both morphological evolutions are intrinsically far from equilibrium features, which were already shown in Kinetic Monte Carlo (KMC) simulations \citep{kaganer2009,toApplSS2021,empting2021,toreis2022,munko2025,lapkinPRL2026} and explained by theoretical approaches \citep{munko2025,lapkinPRL2026}.

Monitoring the evolution of these materials is possible with scanning probe microscopy (SPM) techniques such as atomic force microscopy (AFM) \citep{binnigAFM} and scanning tunneling microscopy (STM) \citep{binnigSTM}, which can provide real-space information on a variety of length scales.
Alternatively, elastic scattering methods have the advantage of being non-invasive and more readily implemented in situ, such as X-ray reflectometry (XRR) \citep{alsnielsenbook} and small-angle X-ray scattering (SAXS) \citep{daillantbook}.
Both classes of techniques have already been separately used to show the rough-to-smooth growth mode \citep{kaganer2009,toApplSS2021}.
In a recent study of the initial stages of $\text{C}_{60}$ film deposition, a discrepancy between the results obtained with different techniques was shown: the AFM images in tapping mode confirmed island growth, coalescence, and formation of a continuous film, with roughness peaks of $1\text{--}1.6$~nm during the coalescence, while XRR revealed a sharp roughness peak of $\approx3$~nm in the same thickness range \citep{lapkinPRL2026}.
The difference between these maximum roughness values is far greater than the data uncertainties and was attributed to the convolution of the AFM tip geometry with the surface topography.

The accuracy in the description of the film geometry is important to warrant the desired physical and chemical properties in applications \citep{Sadewasser2003,Zerweck2005,onoda2021}; for fundamental studies of thin film growth, it is necessary for the correct estimation of microscopic parameters such as the activation energies of surface processes.
For this reason, some works have already analyzed the effects of probe size and probe-surface interaction in SPM \citep{Bonaccurso2006,Johann2010,Lechenault,alvesNPJ2016,amorim2023,xuEFM1995,amorim2025,collinson2021}.
In several cases, the interactions of these probes with self-affine surfaces (i.e., rough surfaces whose fluctuations are distributed in broad wavelength ranges) were considered.
Lechenault et al \citep{Lechenault} showed that probes with parabolic tips changed the height-height correlations at small length scales, but had a small effect on large length scale fluctuations.
Using AFM to scan a fracture surface in tapping mode and using data on the tip geometry of that device, these authors determined numerically the most probable surface profile of that fracture.
Alves \textit{et al.} \citep{alvesNPJ2016} have used KMC simulations and iron electrodeposition to show that probes with wide tips are unable to obtain a full resolution of deep valleys or sloping borders of plateaus, which leads to distortions of height, local roughness, and extremal height distributions.
Recently, Amorim et al \citep{amorim2023} considered self-affine surfaces generated by a growth model and simulated electrostatic force microscopy (EFM) \citep{xuEFM1995} with a cylindrical probe in non-contact mode.
The measured height profiles were obtained from the positions with a constant electric field at the tip and were shown to be similar to the profiles obtained by averaging the original heights inside windows with the same radius as the probe.
This averaging also reduces short-wavelength fluctuations \citep{amorim2025}.
In the context of polymer systems, there are studies of the interactions between their surfaces and AFM tips which help to determine morphological and mechanical properties \citep{collinson2021}.

To our knowledge, no investigation of the effect of probe size in SPM of ultrathin films that follow the rough-to-smooth growth mode has been performed yet, but this is an important step for the accurate characterization of the structure of these materials and may have an impact on their applications.
For this reason, this investigation is performed here with KMC simulations of thin films growing in that mode and considering different models for the interactions between the film surfaces and the SPM probes, with a focus on AFM.
For modeling the non-contact mode, we rely on the conclusion of the recent EFM study \citep{amorim2023} that the measured surface height is approximately equal to the average of the surface heights in a region of the same width as the tip.
This avoids the calculation of forces between the probe and the film surface in 3D space (which would be far from our current computational capabilities) and allows the interpretation of the results in the light of a geometric model where the coalescing islands are treated as a regular array of enlarging cuboids \citep{lapkinPRL2026}.
For modeling contact modes, we consider cylindrical probes with flat apex and pyramidal (or conical) probes with sharp tips.
The probe-surface contact is represented by a hard-core interaction, which means that tip-surface distances may be as low as the diameters of the deposited molecules.
The different behaviors in the different models of probe-surface interaction are also analytically explained through a geometric model.
The recently observed decrease in the roughness peak of $\text{C}_{60}$ films is addressed, with the displacement in the peak of a sample suggesting that ex-situ evolution may also play a role.

\section{Models and Methods}
\label{secMod}

\subsection{Deposition Model}
\label{depositionmodel}

We consider the extended version of the Clarke-Vvedensky model \citep{cv} introduced in Ref. \citep{toApplSS2021} to investigate the growth of semiconductor films on weakly interacting substrates.
In these conditions, the molecules on the substrate diffuse faster than the molecules on the surface of the deposited film (upper layers of the deposit).
We perform the growth on a simple cubic lattice in which the lattice spacing $a=1$ is taken as the length unit and in which a flat substrate is located at $z=0$. 
A deposited atom or molecule occupies a single site $z\geq1$ and can only aggregate on the substrate or on previously deposited particles, so that the deposit has no vacancy (solid-on-solid condition).
For simplicity, the deposited species are hereafter termed adatoms.
We denote as $h(x,y)$ the height of the topmost position of the column of adatoms piled up on a substrate position $(x,y)$.

There is a particle flux $F$ per substrate site per unit time. 
Therefore, in a deposition event, a position $(x,y)$ is randomly selected and the height at that point increases as $h(x,y) \rightarrow h(x,y)+1$ to represent the adsorption of the incoming species.

The topmost adatoms can diffuse by hopping to nearest neighbor (NN) columns respecting the solid-on-solid condition.
Intralayer and interlayer hops are distinguished as follows:

\par\noindent A) Intralayer (in-plane) hops:

Adatoms laying on the substrate and having $n$ lateral neighbors ($0\leq n\leq3$) attempt to hop to the four neighboring columns ($x\pm1$ or $y\pm1$) on the same plane with rate
\begin{equation}
D_1 = D_S \epsilon^n \quad , \quad D_S = \nu e^{-E_S/(k_B T)} ,
\label{defD1}
\end{equation}
where $\nu$ is a constant frequency, $E_S>0$ represents the interaction with the substrate, $k_B$ is Boltzmann's constant, $T$ is the substrate temperature, and
\begin{equation}
\epsilon \equiv e^{-E_B/(k_B T)}
\label{defepsilon}
\end{equation}
is a detachment probability, where the bond energy $E_B>0$ represents the interaction with each lateral NN.

To account for the different properties of the substrate and of the deposited material, an adatom laying on a film layer (i.e., adatoms with $z \geq 2$) can hop to NN sites on the same layer with rate
\begin{equation}
D_2 = D_F \epsilon^n \quad , \quad D_F = \nu e^{-E_F/(k_B T)} .
\label{defD2}
\end{equation}
Assuming that film-substrate interactions are weaker than film-film interactions, we consider $E_F>E_S$ and, consequently, $D_F<D_S$; i.e., the hopping rates of atoms with the same number $n$ of NNs are larger on the substrate than on the upper layers of the deposit.
Observe that the lateral interactions represented by the energy $E_B$ [Eq. (\ref{defepsilon})] are assumed to be the same for adatoms on the substrate and on upper layers.

\par\noindent B) Interlayer (upward or downward) hops:

When an adatom attempts to hop between different layers (i.e. different $z$), it has to cross a step edge and may experience an additional energy barrier $E_{ES}$, which is known as the Ehrlich-Schwoebel (ES) barrier \citep{ehrlich,ES}.
For a hop of unit height difference, the hopping rate is reduced (in comparison with the intralayer rate) by a factor
\begin{equation}
P \equiv e^{-E_{ES}/(k_B T)} \quad \left( \Delta z=1\right) .
\label{defP}
\end{equation}
When the height difference is $\Delta z>1$, a hop between two neighboring columns represents the crossing of a step edge plus the adatom diffusion along a vertical facet separating those columns; this is a feature of the simple cubic geometry discussed in Ref. \protect\cite{toreis2022}.
During that off-plane diffusion, the adatom may return to its initial position before reaching the other layer, effectively yielding an additional decrease in this hopping rate.
However, the off-plane diffusion does not respect the solid-on-solid condition, so it has to be represented by an effective reduction $P_{hop}$ of the hopping rate between the tops of the neighboring columns.
Assuming that the off-plane diffusion is restricted to the vertical direction and is much faster than the horizontal hops, a simple expression for this reduction factor (in comparison with the intralayer rate) is \citep{lealJPCM}
\begin{equation}
P_{hop} = \frac{P}{1 + P(\Delta z - 1)} .
\label{defPhop}
\end{equation}
Observe that $P_{hop}=P$ for $\Delta z=1$, as expected, and $P_{hop}<P$ for $\Delta z>1$ due to the effects of the hidden off-plane diffusion (a kinetic barrier in addition to the ES barrier).

Two important dimensionless parameters to characterize the interplay between temperature and deposition flux are the diffusion-to-deposition ratios on terraces, which are defined as
\begin{equation}
R_S = \frac{D_S}{F} = \frac{\nu}{F}e^{-E_S/(k_B T)}, \quad R_F = \frac{D_F}{F} = \frac{\nu}{F}e^{-E_F/(k_B T)} .
\end{equation}
The other dimensionless model parameters are the detachment rate $\epsilon$ [Eq. (\ref{defepsilon})] and the ES barrier factor $P$.

\subsection{Models of probe-surface interactions}
\label{probemodels}

Three models are proposed to simulate the effects of finite sizes of the SPM probes on the reported height profiles.
Hereafter, the term original height profile [$\left\{h(x,y)\right\}$] is used to denote a profile obtained in the KMC simulations, while the term transformed profile denotes the profile calculated with some model using an original profile.
Model I is designed to represent scanning in non contact mode, while models II and III represent effects in contact mode.

\par
\begin{figure}[!t]
\includegraphics[width=0.5\linewidth]{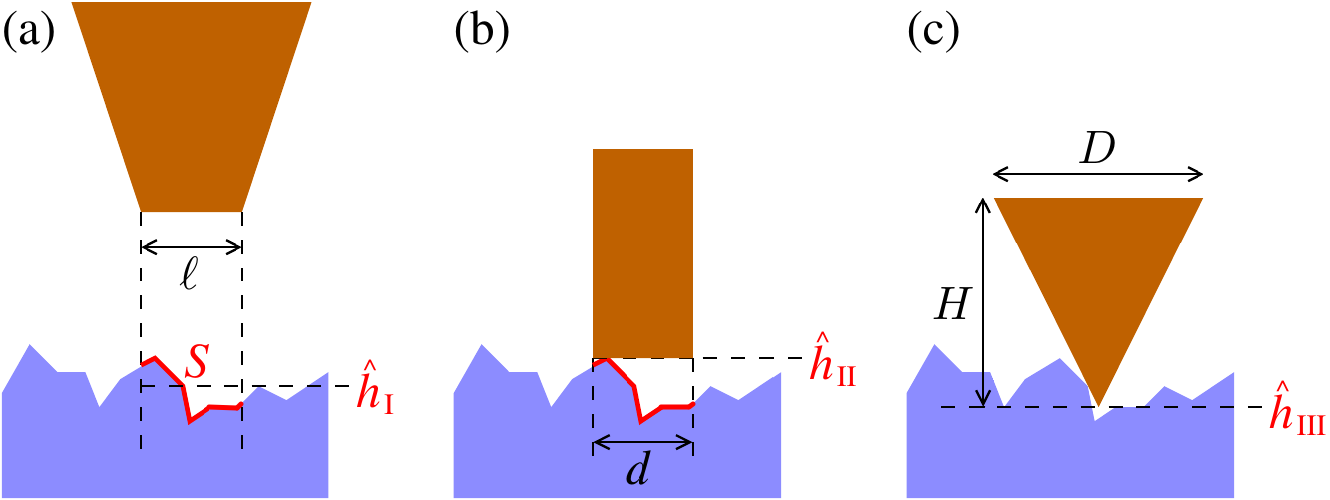}
\caption{Schematic lateral view of a film (blue) and the AFM probe tips (brown) considered in (a) model I, (b) model II, and (c) model III, all them centered at the same position of the film surface.
The measured heights $\hat{h}$ in each case are indicated by the horizontal dashed lines. The region $S$ sampled by the tip is marked in red in (a) and (b), while in (c) it corresponds to the diameter $D$ of the tip.}
\label{probes}
\end{figure}

\noindent \textit{I}) The measured height profile is approximated by averaging the original interface heights inside a window of the same width as that of a flat probe tip.
Specifically, for a given probe and an original profile $\left\{h(x,y)\right\}$, the transformed height profile $\left\{\hat{h}_I(x,y)\right\}$ is given by
\begin{equation}
\hat{h}_I(x,y) = \frac{1}{A} \sum_{(x',y') \in S} h(x',y') ,
\label{hI}
\end{equation}
where $S$ is a box centered at $(x,y)$ and $A$ is the number of sites $(x',y')$ inside this box.
To account for different shapes of the probe tips, we consider square boxes of side $\ell$ aligned with the lattice directions ($x$ and $y$) and circular boxes of diameter $d$ centered at lattice sites, in which $A = \ell^2$ and $A \approx \pi d^2/4$, respectively; the square box case is illustrated in a two-dimensional projection in Fig. \ref{probes}(a).

Model I is justified by a recent study of the differences between the actual shapes of lattice-based self-affine interfaces and the shapes obtained in EFM under two conditions \citep{amorim2023}:
(i) the lateral widths of the probes could be much larger than the lengthscales of small fluctuations of the original interfaces, which are on the order of the lattice constant; and
(ii) the distances between the probe tips and the interfaces exceeded the amplitudes of the largest lengthscale fluctuations of the latter, so a non-contact mode was represented.
The measured height profiles were obtained from the tip positions in which the applied field at the tip had a constant value, which demanded the solution of the Laplace equation in the region between the probe and the real surface.
It was shown that the measured height profiles were approximately the same as those obtained by averaging the original interface heights inside a window of the same width as the probe tip.
This averaging is adopted in model I because it is computationally much simpler than solving the Laplace equation, particularly in 3D space.

\par\noindent \textit{II}) To represent the effect of a finite width of the probe tip in contact mode, Lechenault \textit{et. al} \citep{Lechenault} proposed to consider that, at a given position $(x,y)$ where the probe is centered, the probe is slowly displaced towards the surface and stops at its first contact with the surface (i.e. assuming hard-core interactions between the probe and the surface).
Therefore, the transformed height $\hat{h}(x,y)$ is the height of the probe center when it stops.
Parabolic shapes were considered in that study, as suggested for pristine tips of wide use a few decades ago \citep{weisenhorn}.

The first probe considered here is of cylindrical shape with a flat apex, which may be viewed as an approximation to probes built with carbon nanotubes \cite{dai1996,nguyen2001}.
The cylinder axis is parallel to the $z$ direction and is constrained to cross the center of a substrate site.
The first contact of this probe with the surface occurs at the maximal surface height inside a box of the same diameter as the cylinder and centered at the cylinder axis.
The transformed height is then given as
\begin{equation}
\hat{h}_{II}(x,y) = \max_{(x',y') \in S}\left\{h(x',y')\right\} .
\label{hII}
\end{equation}
Again, $S$ may be a circular box with diameter $d$ or a square box with side $\ell$ centered at $(x,y)$; see illustration in Fig. \ref{probes}(b).

\par\noindent \textit{III}) For checking other possible effects of a finite probe size in contact mode, we consider probes with pyramidal and conical shapes, both having very sharp tips.
They are schematically illustrated in Fig. \ref{probes}(c).
The pyramidal probes have a square base of lateral size $D$ and height $H$, so the area of the base is $A = D^2$.
The conical probes have a circular base of diameter $D$ and height $H$, so $A = \pi D^2/4$.

When the probe tip is located at an axis crossing the substrate position $(x,y)$, the probe is slowly displaced towards the surface, stops at the first contact with the surface, and the transformed height is the position $z$ of the tip center at that point.
The first contact may be at the sharp tip or not; see Fig. \ref{probes}(c).
For a probe shape given by the continuous function $g(x,y)$, the transformed height is
\begin{equation}
\hat{h}_{III}(x,y) = \max_{(x',y') \in S}\left\{h(x',y')-g(x-x',y-y')\right\} ,
\label{hIII}
\end{equation}
where the region $S$ of the surface is that enclosed by a vertical projection of the probe base.

An important geometrical parameter to characterize these probes is their aspect ratio defined as
\begin{equation}
{\mathcal R} = \frac{D}{H} .
\label{aspect}
\end{equation}
We will generally consider probes with ${\mathcal R} \leq 1$ (i.e. vertically elongated probes).

\subsection{Quantities of interest}

The film coverage is the number of adatoms per site (or number of deposited layers) and is given as
\begin{equation}
\theta = Ft .
\label{defthera}
\end{equation}

The surface roughness at a given coverage $\theta$ or time $t$ is the standard deviation of the surface heights:
\begin{equation}
W = \left< \left[\overline{\tilde{h}^2} - \bar{\tilde{h}}^2\right]^{1/2} \right> ,
\label{defW}
\end{equation}
where $\left\{ \tilde{h}(x,y)\right\}$ may represent the heights of the simulated surfaces ($\tilde{h}=h$) or of the transformed surfaces ($\tilde{h}=\hat{h}$), the overbars denote spatial averages, and the angular brackets denote an ensemble average.

\subsection{Simulation details}
\label{details}

The main interest of this study is to determine the effects of large probe sizes that scan deposits growing in the rough-to-smooth mode, in which multilayer islands nucleate, grow, and coalesce.
As shown in recent works, this mode is clearly observed when $R_S$ and $R_F$ differ by at least $2$ orders of magnitude and when $P=1$, corresponding to a null ES barrier \citep{toApplSS2021,toreis2022}.
Effects of $\epsilon$ are generally less important.

Guided by these observations, here we consider $P=1$ in all simulations and four sets of the other parameter values, as listed in Table \ref{tab1}.
Observe that $P_{hop}<1$ for interlayer hops with $\Delta z>1$ [Eq. (\ref{defPhop})], i.e. the kinetic barrier due to the hidden off-plane diffusion is still present.

\begin{table}[!t]
\caption{Dimensionless parameters used in the simulations.}
\begin{ruledtabular}
\begin{tabular}{l c c c c c c}
Set          & &      $R_S$               & &      $R_A$           & &      $\epsilon$             \\
\hline
$A$          & &      $10^7$              & &      $10^4$          & &      $10^{-3}$              \\
$B$          & &      $10^8$              & &      $10^5$          & &      $10^{-3}$              \\
$C$          & &  $3.9\times 10^7$        & &  $1.5\times 10^5$    & &  $3.9\times 10^{-3}$        \\
$D$          & &  $2\times 10^8$          & &  $5.7\times 10^5$    & &  $2.8\times 10^{-3}$        \\
\end{tabular}
\end{ruledtabular}
\label{tab1}
\end{table}

The simulations are performed in cells of a large lateral size $L=2048$ with periodic boundaries in the $x$ and $y$ directions.
As discussed in previous studies  \citep{toApplSS2021,toreis2022}, this avoids effects of the finite cell size when the coverages are relatively small, which is the case of the simulations here.
Moreover, such large cells allow that many representative microscopic environments (e.g. many islands with different shapes) are represented in a single sample.
For this reason, a single sample is representative of an infinitely large substrate; see Sec. \ref{originalsurfaces}.

The simulated probes of models I and II have tip widths (or diameters) ranging from $3$ to $33$ lattice units.
For inorganic films with $a\sim0.3$~nm, these values are in the range $1\text{--}10$~nm, while for films of organic molecules with $a\sim1$~nm that range is $3\text{--}30$~nm.
These are values consistent with real AFM probes.
The base sizes $D$ of the pyramidal and conical probes of model III are of the same order of magnitude.

\section{Results and Discussion}
\label{results}

\subsection{The original film surfaces}
\label{originalsurfaces}

\begin{figure}[!ht]
\includegraphics[width=7cm]{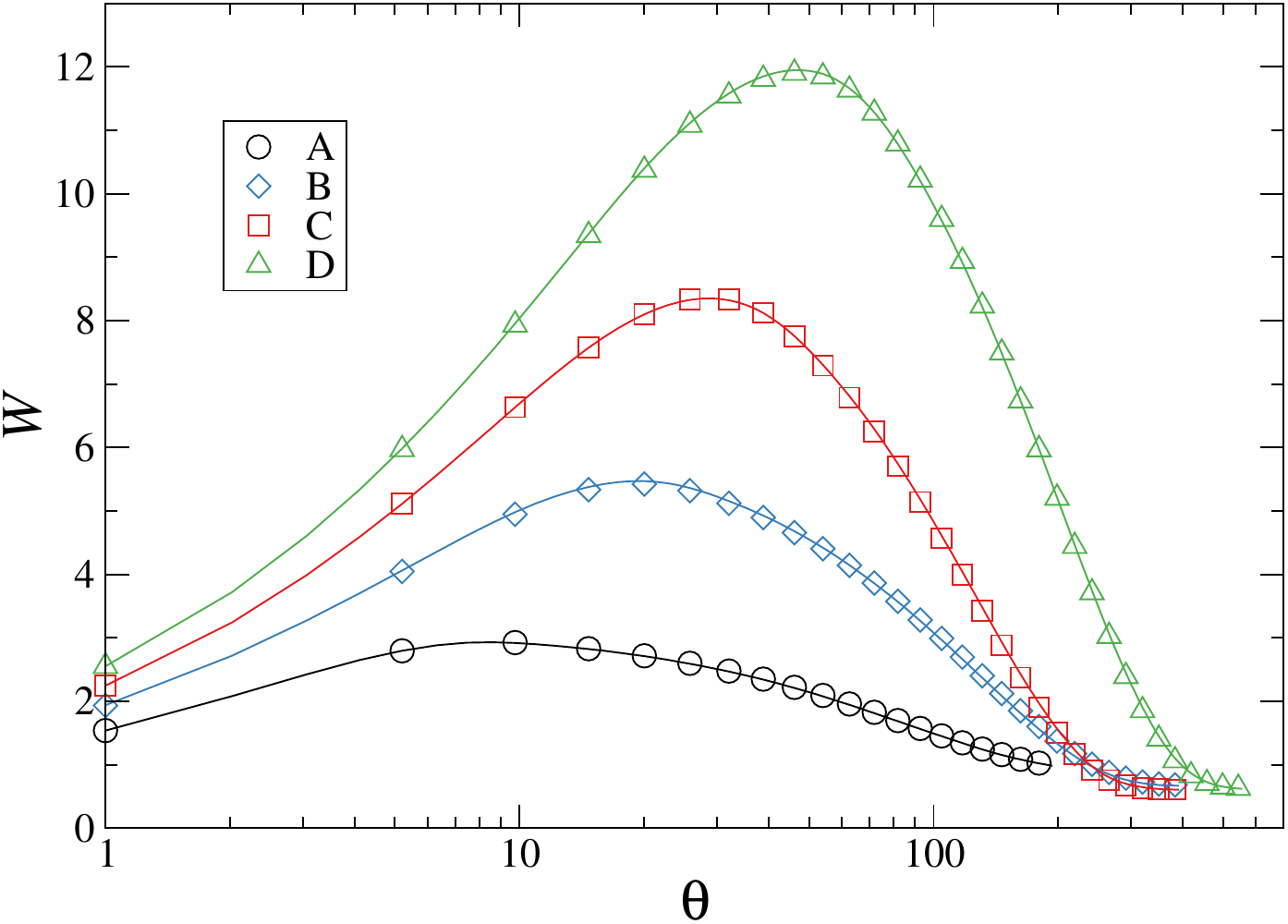}
\caption{Roughness as a function of coverage (in lattice units) for films grown with the four parameter sets. For each set, symbols represent the data collected from the particular realization used in the following analyses, while solid lines show averages over ten independent realizations.
}
\label{comparison}
\end{figure}

Figure \ref{comparison} compares the evolution of the roughness $W$ for a particular run (symbols) with the average over ten independent runs (lines). Thanks to the very large system size ($L=2048$), the curve obtained from a single realization overlaps with the ten-realization average for all parameter sets.
This indicates that a single realization contains all representative microscopic environments of the growth process and, consequently, provides a good approximation of the ensemble average defined in Eq. (\ref{defW}).
For this reason, the results shown in the following subsections correspond to this single representative realization for each parameter set. From an experimental standpoint, this is important because the probe-size effects reported here should also be representative of an individual sample, provided that the measured region is representative of the surface as a whole.

Figure \ref{comparison} also confirms that the four parameter sets lead to film growth in the rough-to-smooth mode \citep{lapkinPRL2026}.
The maximal roughness $W_{max}$ is obtained at a thickness denoted as $\theta_{max}$ and we observe that $W_{max}$ increases as $\theta_{max}$ increases;
the former varies from $\approx3$ to $\approx12$, while the latter varies from $\approx8$ to $\approx 47$ (in lattice units).
As explained in a previous study, $99\%$ of the substrate is covered when the thickness $\theta$ is larger than, but of the same order of magnitude as $\theta_{max}$ \citep{toreis2022}.
Thus, the maximal roughness characterizes the island coalescence \citep{toreis2022}.

The film growth after island coalescence depends on the parameters $R$ and $\epsilon$, but not on the film-substrate interaction.
These films have small roughness, typically in the range $0.3\text{--}0.8$ lattice units.
These values are consistent with those of homoepitaxially grown films with parameters $R$ and $\epsilon$ of the same order as those of sets A-D and thicknesses of ${10}^2\text{--}{10}^3$ layers \citep{carrascoPRE2023}.
Although these films are generally not growing in LBL mode, their surfaces are mostly formed by terraces in $3\text{--}4$ consecutive heights, which explains their small roughness.

\subsection{General features of the transformed surfaces}
\label{general}

\begin{figure}[!ht]
\includegraphics[width=0.9\textwidth]{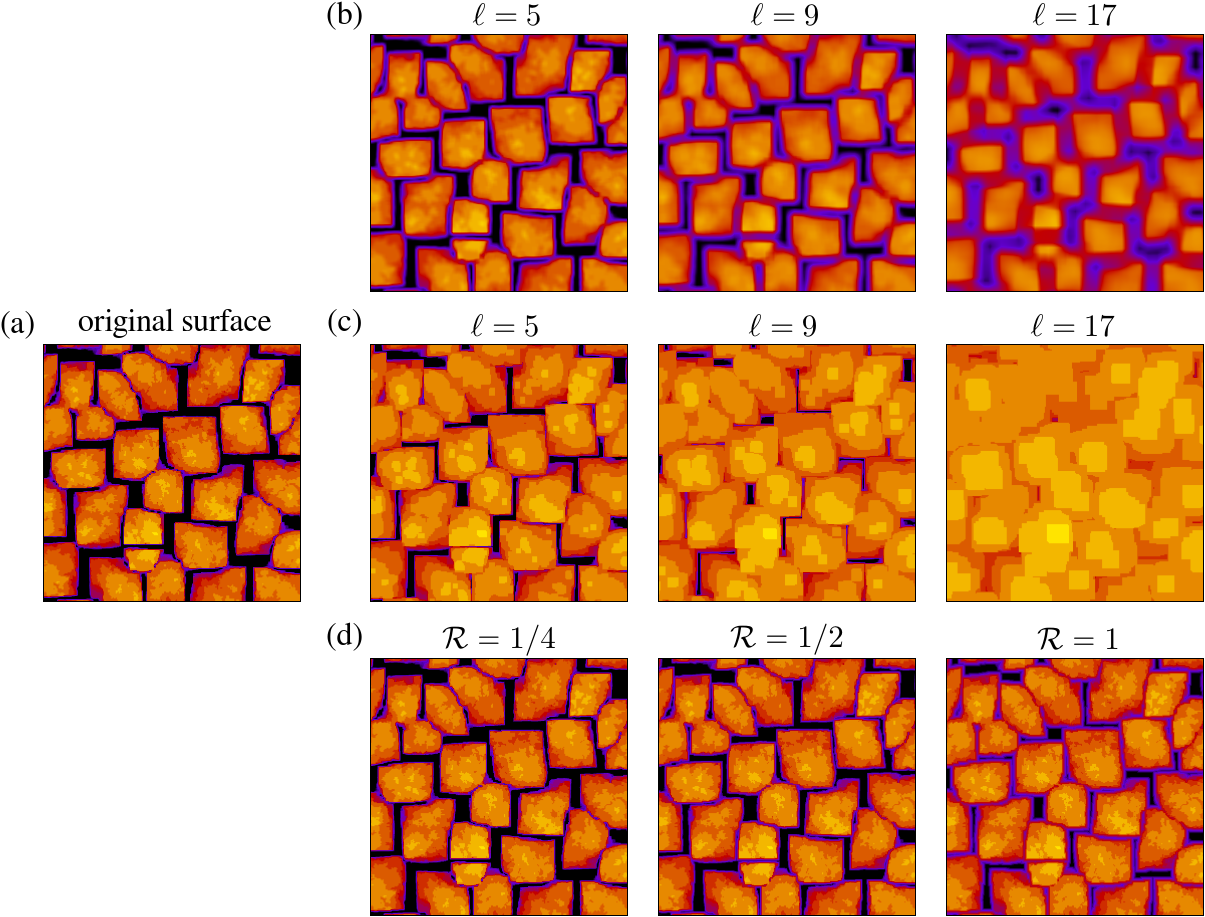}
\caption{Snapshots of size $256\times256$ taken from: (a) the original surface grown with set B at coverage $\theta = 9.4574$; (b), (c), and (d) the transformed surfaces obtained from (a) using models I, II, and III, respectively.
The results in (b) and (c) are for square boxes with the indicated sizes.
Case (d) is for pyramidal tips with base $D=33$ and the indicated aspect ratios.}
\label{snapshots}
\end{figure}

Figure \ref{snapshots}(a) shows a snapshot of part of a deposit grown with set B near the point where the roughness reaches its maximal value; see Fig. \ref{comparison}.
The deposit has islands of approximately square shapes with lateral sizes $\sim40$ (in lattice units) and separated by narrower gaps where the substrate is exposed.
The top parts of those islands are irregular mounds with small average slopes and there are abrupt cliffs separating islands and gaps.

The mean height transformation (model I) is applied in Fig. \ref{snapshots}(b) with square tips of three lateral sizes.
The island borders become diffuse because the original cliffs at those borders are smoothened; for this reason, the gaps between the islands become much narrower.
The island tops get smoother, but their maximal heights weakly change.
This is particularly clear for the widest probe tips with $\ell=17$, which is nearly half of the typical island width and larger than most gap widths.
Overall, the transformed surface appears to be less rough.

Cylindrical probes in contact mode (model II) are considered in Fig. \ref{snapshots}(c).
A more drastic suppression of the gaps is observed, which is reasonable for a model that returns the largest height near the cliffs that separate islands and gaps.
For the same reason, the topmost islands are larger than those of the original image.
When the largest tip width is considered ($\ell=17$), most of the surface is formed by terraces of $3\text{--}4$ consecutive heights, so it is much smoother than the original surface.

The results of scanning with the pyramidal probes in contact mode (model III) are shown in Fig. \ref{snapshots}(d) for three aspect ratios ${\mathcal R}$ and the same base diameter $D=33$.
For a small ratio ${\mathcal R}=0.25$ (i.e. a sharp probe), the lateral side of the probe seldom touches the surface before the tip; this lateral contact is restricted to a small number of points near the island borders.
For this reason, the transformed image is visually similar to the original one.
Significant discrepancies between the original and the transformed images are observed only for the highest aspect ratio ${\mathcal R}=1$, which smoothens the island borders; however, no effect is visually observed in the island tops and in the middle of the gaps.

Figures S1-S3 of the Electronic Supplement show snapshots of parts of the deposits grown with the other sets (A, C, and D) and snapshots of the corresponding transformed surfaces with models I, II, and III.
The results are similar to those of Figs. \ref{snapshots}(a)-(d), although the island size vary substantially.

\subsection{Transformation with the mean height method (model I)}
\label{secResWHM}

\begin{figure*}[!ht]
\includegraphics[width=0.24\linewidth]{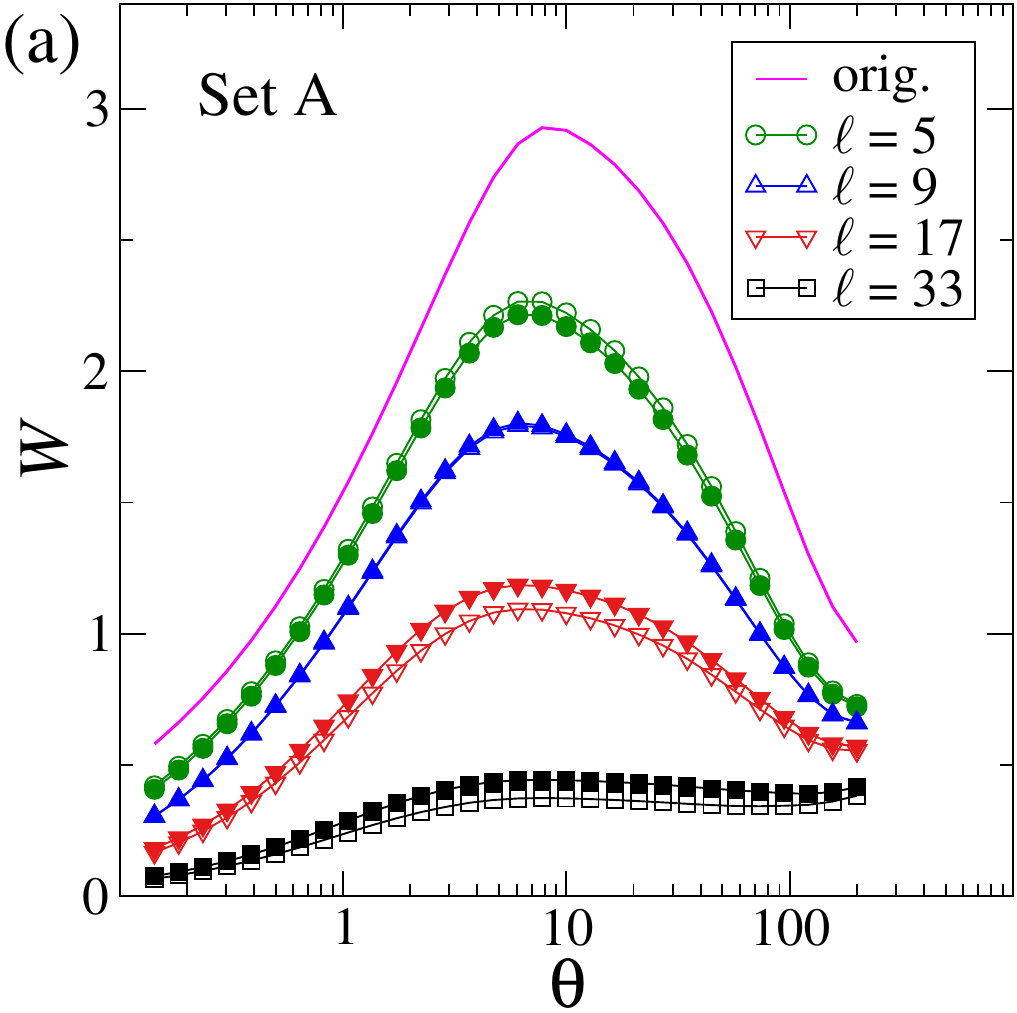}
\includegraphics[width=0.24\linewidth]{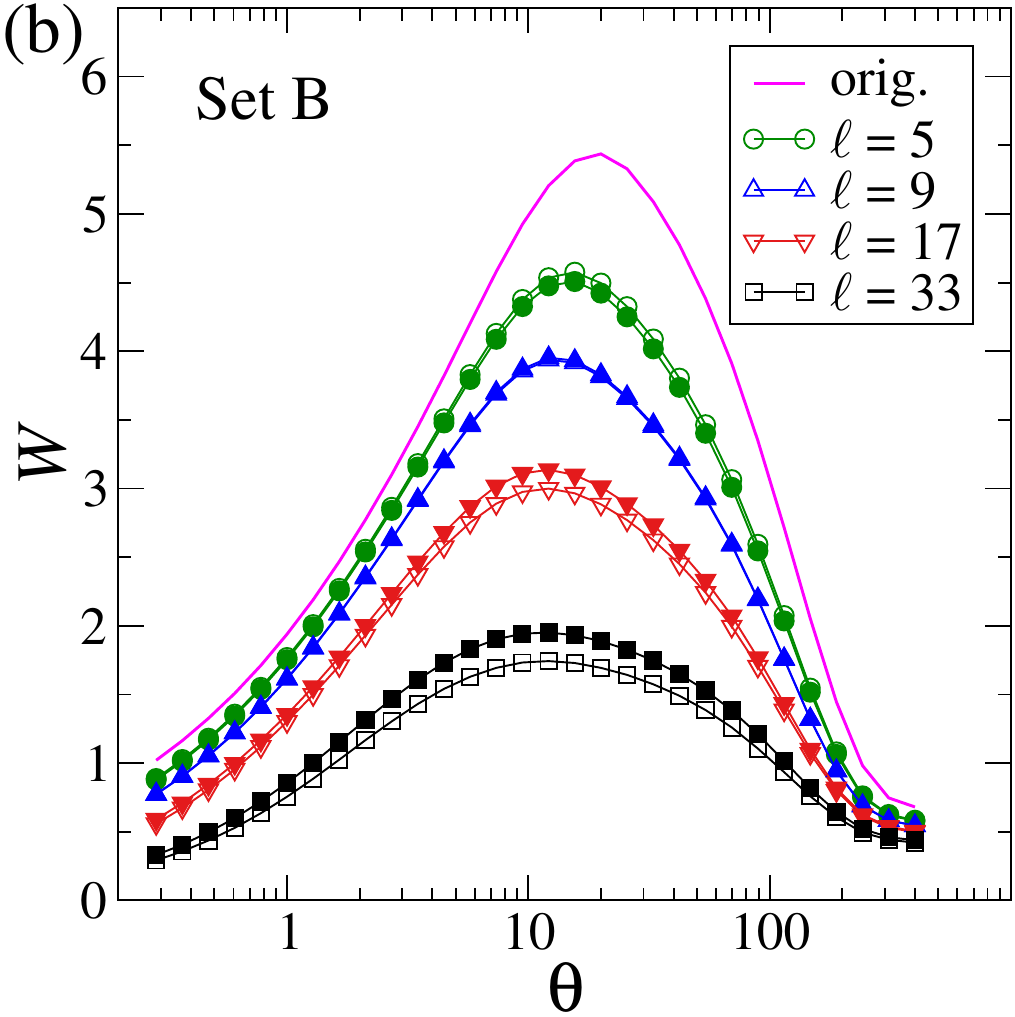}
\includegraphics[width=0.24\linewidth]{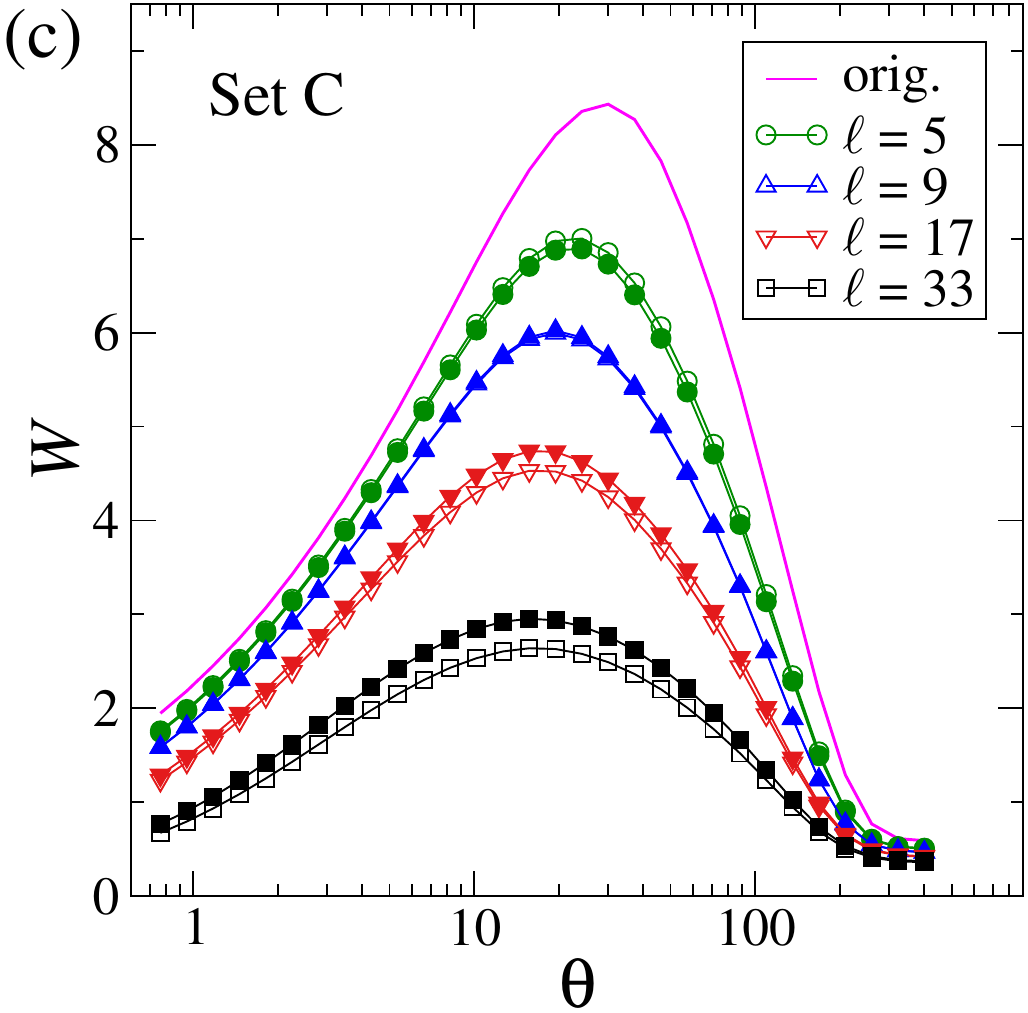}
\includegraphics[width=0.24\linewidth]{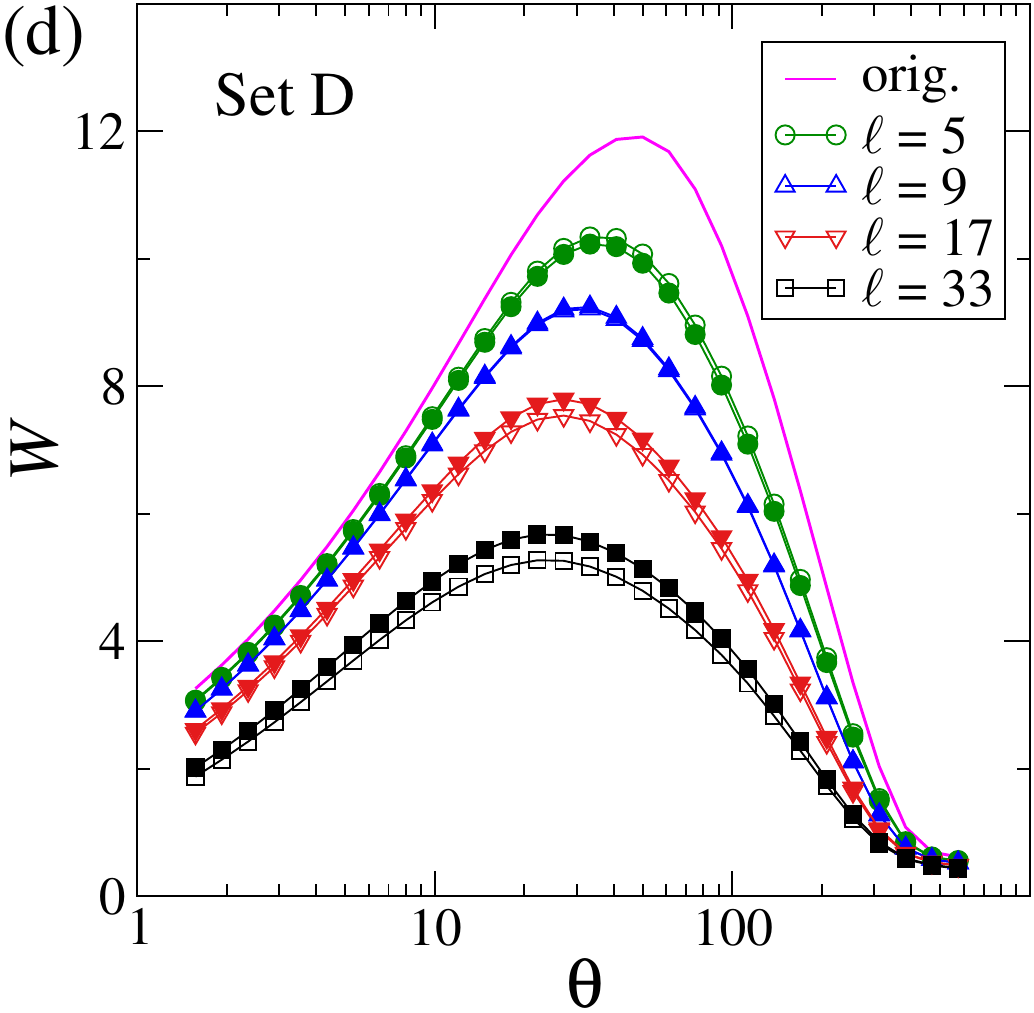}
\caption{Roughness as a function of the coverage for transformed surfaces with model I, in which the average height is calculated inside square windows of lateral size $\ell$ (open symbols) or circular windows of diameter $d=\ell+2$ (full symbols). The set of growth parameters and the values of $\ell$ are indicated in each panel. The curves without symbols are the roughness evolutions of the original surfaces.}
\label{roughnessI}
\end{figure*}

Figures \ref{roughnessI}(a)-(d) show the evolution of the roughness $W$ of surfaces grown with the four parameter sets of Table \ref{tab1} and of the surfaces transformed with the mean height method (model I).
The results for square boxes of lateral size $\ell$ are very close to those for circular boxes of diameter $d=\ell+2$, so only the value of $\ell$ is indicated for both tip shapes.

The thickness $\theta_{max}$ in which the roughness peak occurs is generally preserved in the evolution of the transformed surfaces.
The exception is the case of the largest $\ell$ ($=33$) in films grown with the parameter set A, but this is a case in which $\ell$ is close to the average island size and exceeds the average gap width; see Fig. S1 of the Electronic Supplement.
As $\ell$ increases, one observes that $W$ decreases for a given coverage, which is the smoothening effect suggested by inspection of the transformed images.
This indicates a trend of monotonically decrease of $W$ with $\ell$; observe that $W\rightarrow 0$ as $\ell \rightarrow L$ because the averaging includes all points of the original surface in this limit.

\begin{figure}[!ht]
\includegraphics[width=0.24\linewidth]{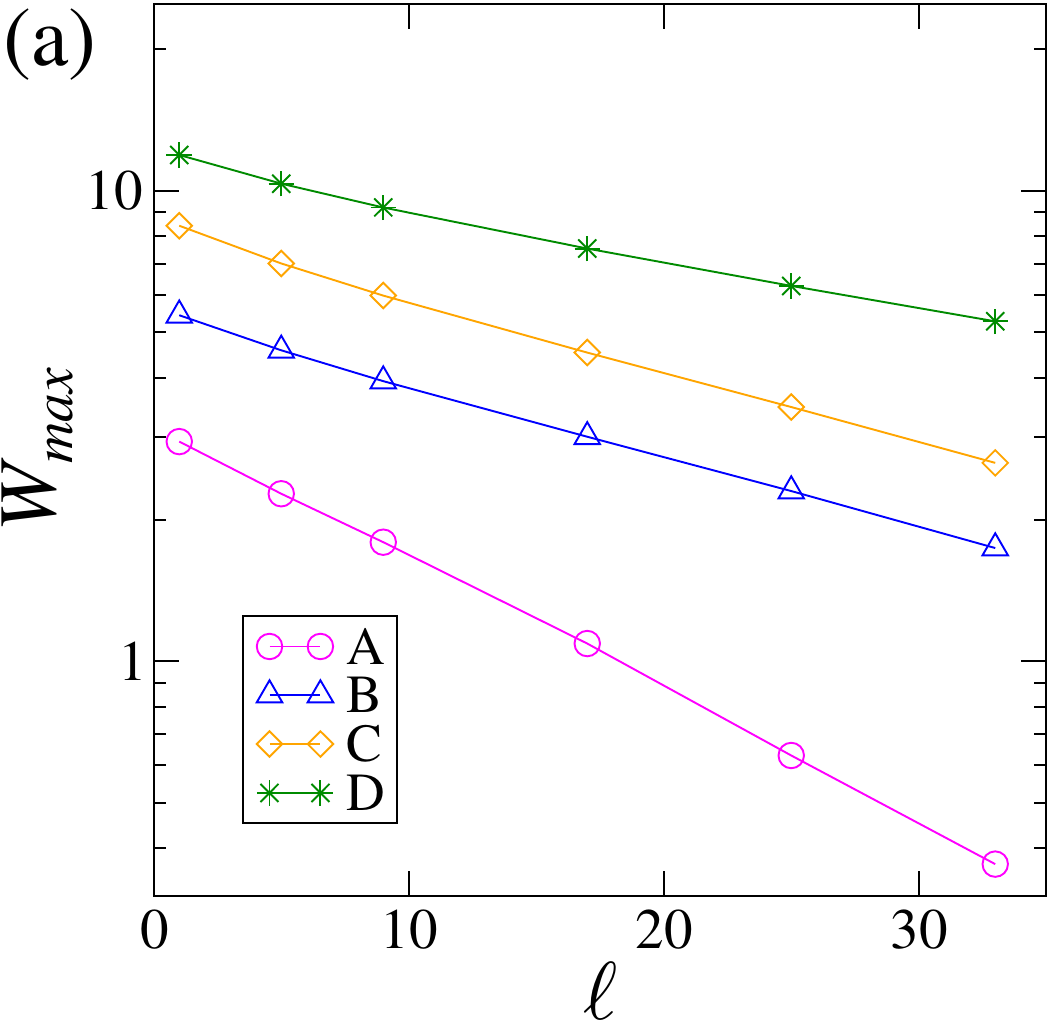}
\includegraphics[width=0.24\linewidth]{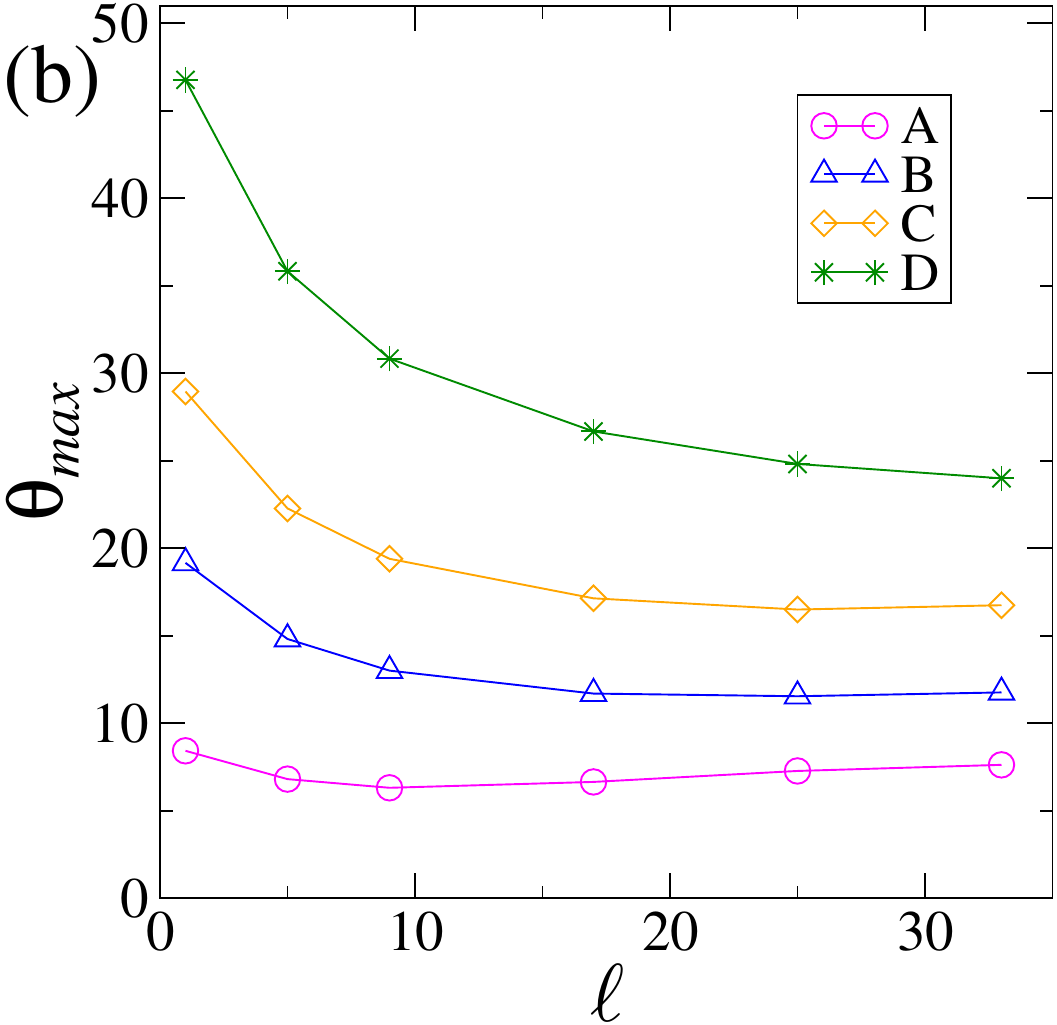}
\caption{(a) Maximal roughness as a function of the width $\ell$ of the square boxes used in the surface transformation with model I. (b) The corresponding thicknesses at which the roughness is maximal as a function of $\ell$.
Data for different sets of parameters are shown, as indicated in the legends.}
\label{maximoI}
\end{figure}

Correspondingly, the maximal roughness $W_{max}$ of the transformed surfaces shows a considerable decrease with $\ell$.
Figure \ref{maximoI}(a) shows $W_{max}$ for square boxes as a function of $\ell$ in linear-log scale, which indicates an effectively exponential decay in the studied range.
The characteristic lengths of these decays are: $\ell^*\approx 15$ for set A, $\ell^*\approx 28$ for sets B and C, and $\ell^*\approx 40$ for set D; only the latter is larger than the maximal box width $\ell=33$ studied here.
On the other hand, after a fast decrease for $\ell \lesssim 10$, the thicknesses $\theta_{max}$ in which the maximal roughness is observed only have a mild variation with $\ell$, as shown in Figure \ref{maximoI}(b).

\subsection{Transformation with cylindrical probes (model II)}
\label{secResWHmax}

Now we consider the situation where the films are scanned with probes consisting of vertical cylinders with flat apex of square or circular forms and the contact mode is modeled by intersections between the probe tip and the film surface.
This simply returns the maximal height within each square box (of lateral size $\ell$) or circular box (of diameter $d$) below each position.

\begin{figure*}[!ht]
\includegraphics[width=0.24\linewidth]{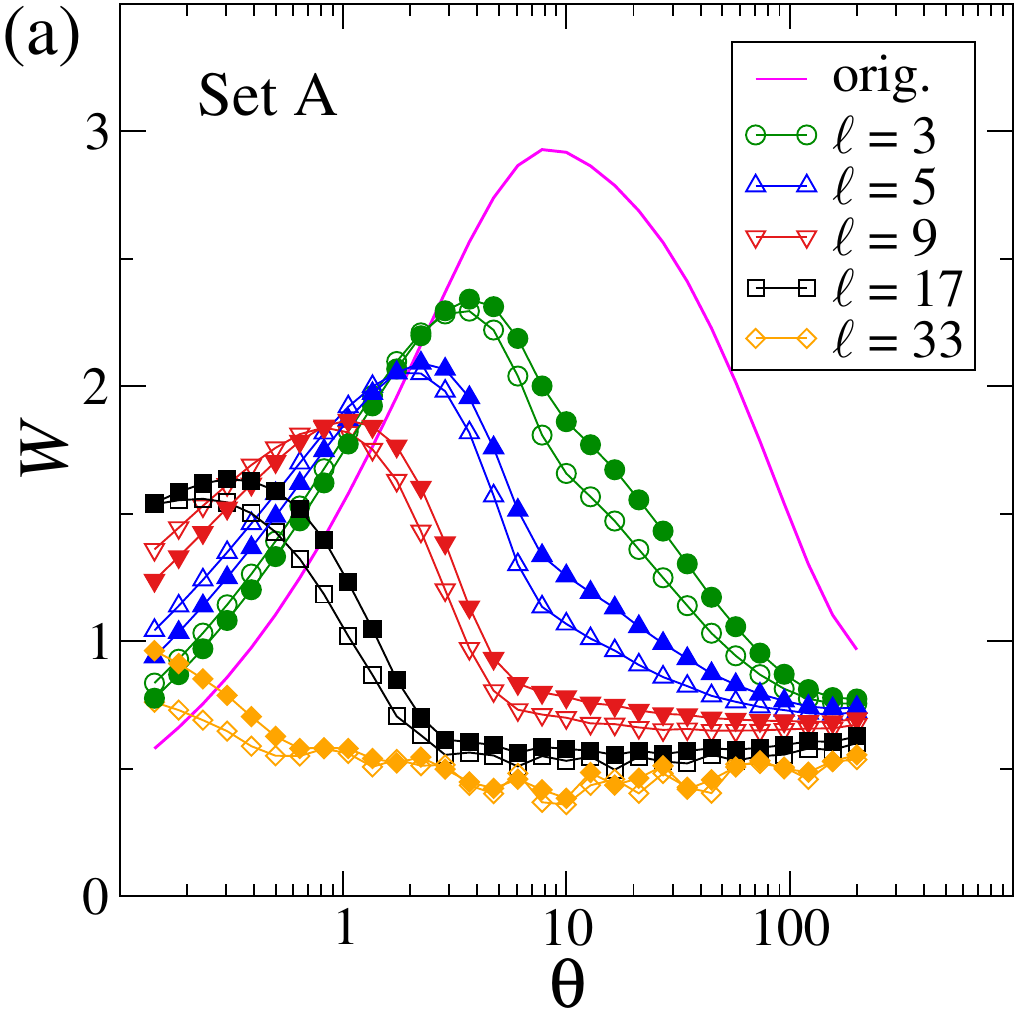}
\includegraphics[width=0.24\linewidth]{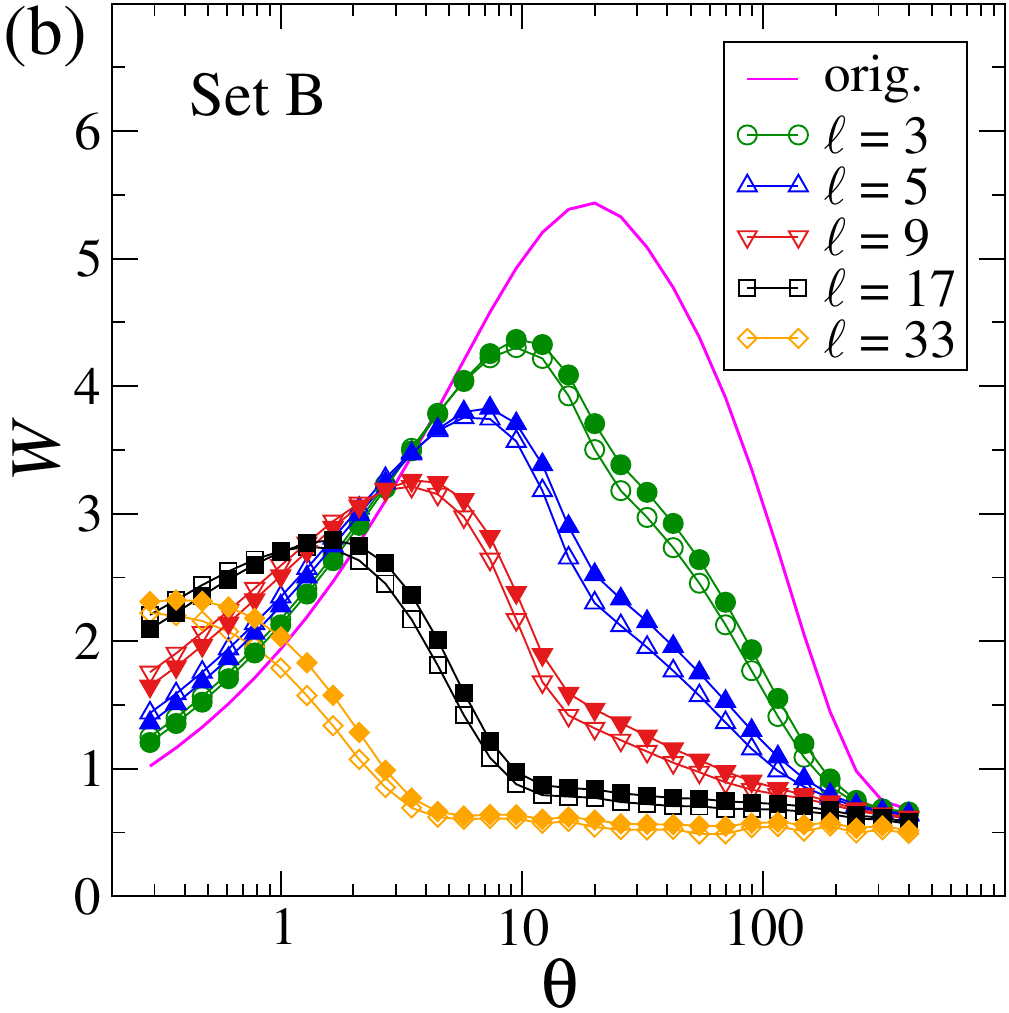}
\includegraphics[width=0.24\linewidth]{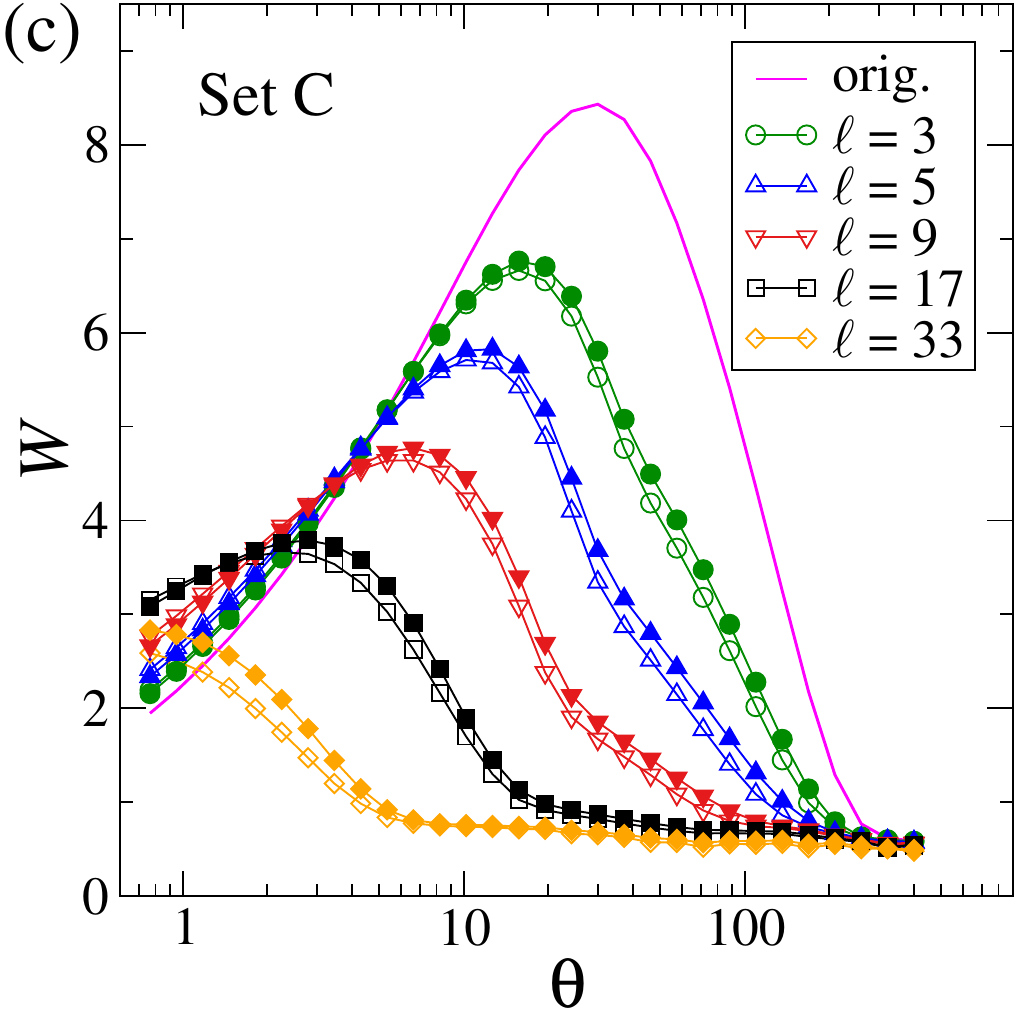}
\includegraphics[width=0.24\linewidth]{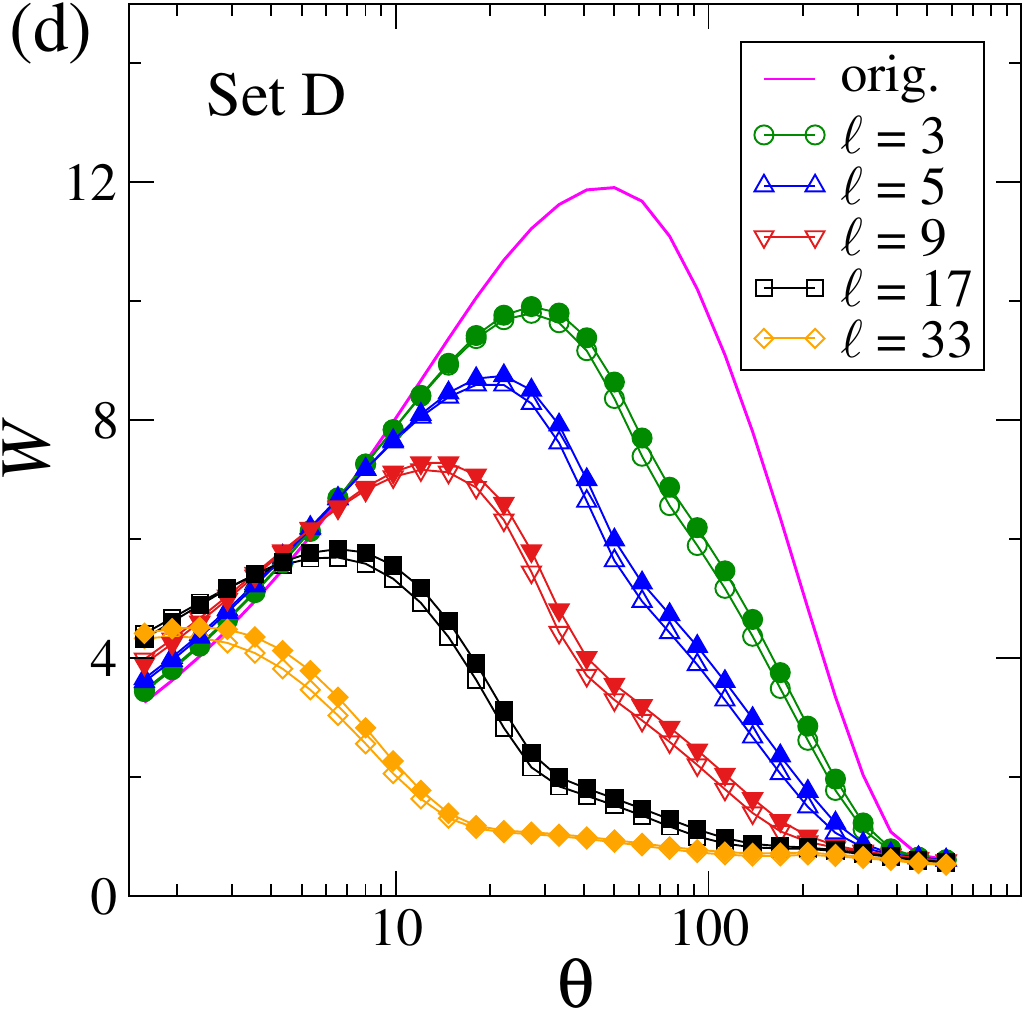}
\caption{Roughness as a function of the thickness for transformed surfaces with model II, in which the maximal height is calculated inside square windows of lateral size $\ell$ (open symbols) or circular windows of diameter $d = \ell$ (full symbols). The set of growth parameters and the values of $\ell$ are indicated in each panel. The curves without symbols are the roughness evolutions of the original surfaces.}
\label{roughnessII}
\end{figure*}

Figures \ref{roughnessII}(a)-(d) present the evolution of the roughness $W$ of surfaces grown with the four parameter sets of Table \ref{tab1} and of the surfaces transformed with this maximal height method (model II).
The roughnesses for square boxes and circular boxes with $\ell=d$ have approximately the same value, indicating a weak effect of the tip geometry.
Similarly to model I of surface transformation, a smoothening of the surface is observed here as $\ell$ increases; this is also consistent with the visual observation of the snapshots of the transformed surfaces [Fig. \ref{snapshots}(c)].
However, in contrast with model I, the thickness $\theta_{max}$ displays a fast decrease with $\ell$ or $d$.
For the largest probe sizes, $\ell=d=33$, the roughness peak may disappear or be located at $\theta_{max}<1$, which implies a monotonic decrease of the roughness from early growth times to the largest thicknesses ($\theta=400\text{--}600$).
In an experiment with such a scanning probe, the rough-to-smooth mode would not be revealed by the measured roughness.

Another interesting feature of model II is the increase of the roughness with the probe size at short times.
This occurs when the islands are narrow and sparsely distributed, so the probes that detect the maximal height inside large areas will artificially increase the island widths; this leads to larger roughness.
However, the maximal island heights do not change, so the roughness peak remains in most cases.

\begin{figure}[!ht]
\includegraphics[width=0.24\linewidth]{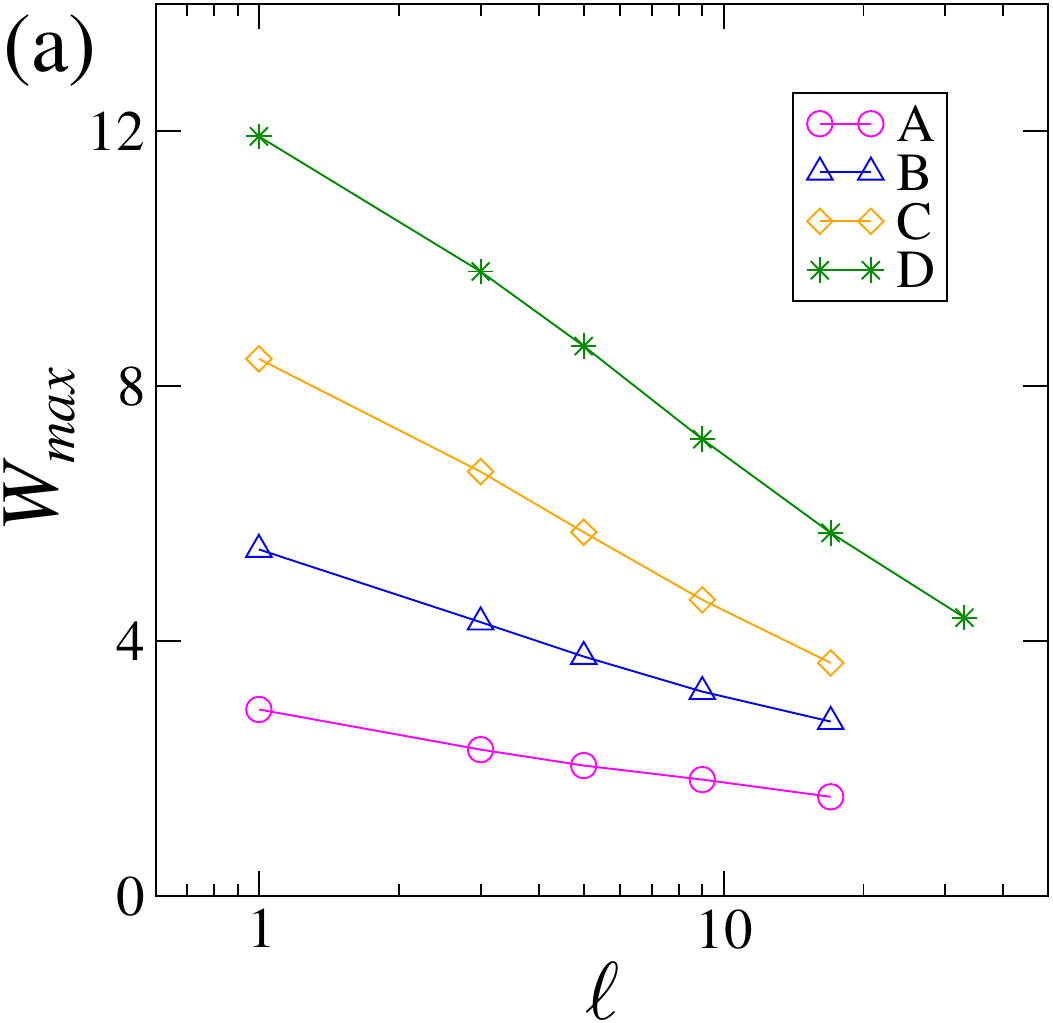}
\includegraphics[width=0.24\linewidth]{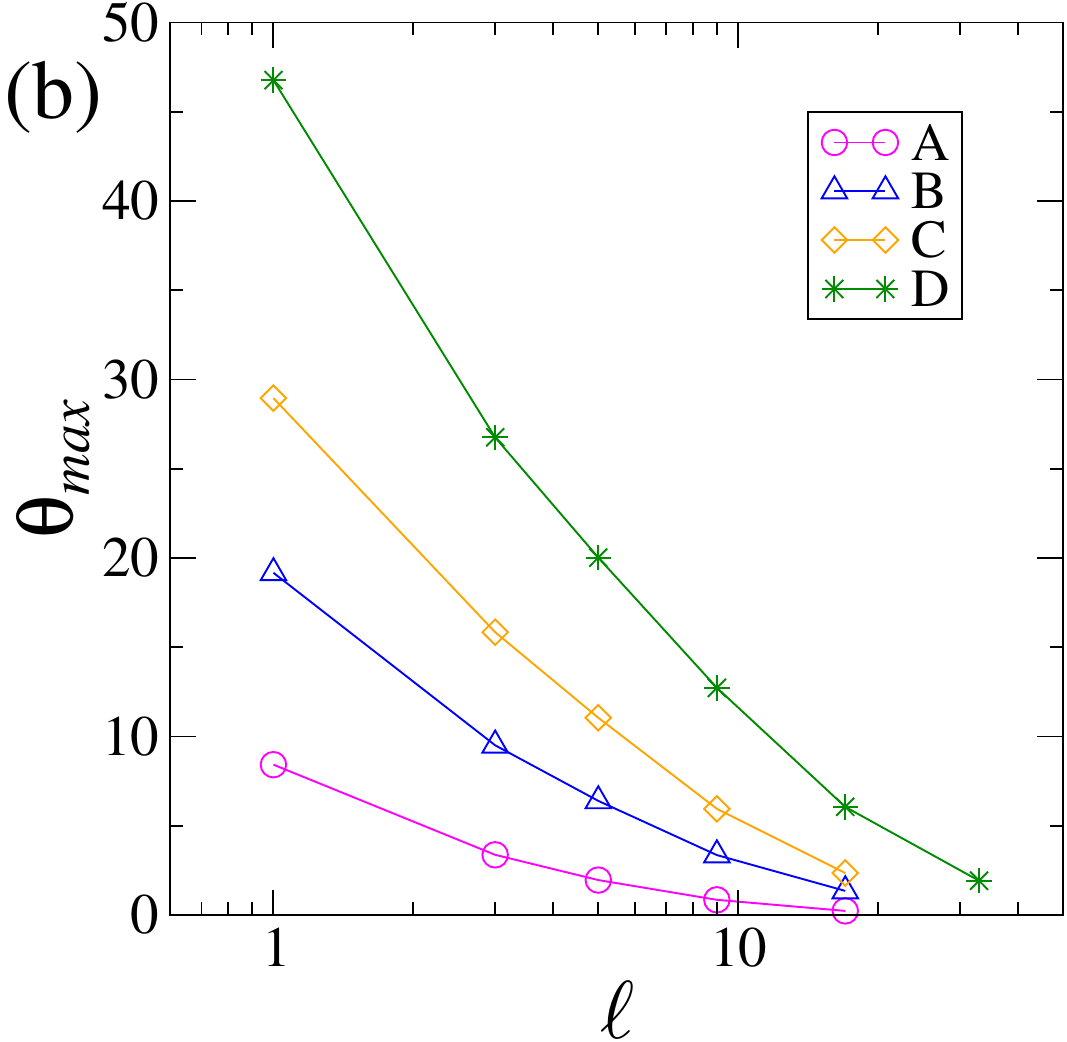}
\caption{(a) Maximal roughness as a function of the width $\ell$ of the square boxes used in the surface transformation with model II. (b) The corresponding thicknesses at which the roughness is maximal versus $\ell$. Data for different sets of parameters are shown, as indicated by the legends.}
\label{maximoII}
\end{figure}

Figure \ref{maximoII}(a) shows that the roughness peak decreases as $W_{max} \approx b_1-b_2\log(\ell)$ (with $b_1$ and $b_2$ constants) in the studied range of box width.
This differs from the approximately exponential decay of model I [Fig. \ref{maximoI}(a)], but the changes in $W_{max}$ are of the same order of magnitude in the two models.
Figure \ref{maximoII}(b) confirms the rapid decay of $\theta_{max}$ with $\ell$.

\subsection{Transformation with pyramidal probes (model III)}
\label{secResWcylind}

\begin{figure*}[!ht]
\includegraphics[width=0.22\linewidth]{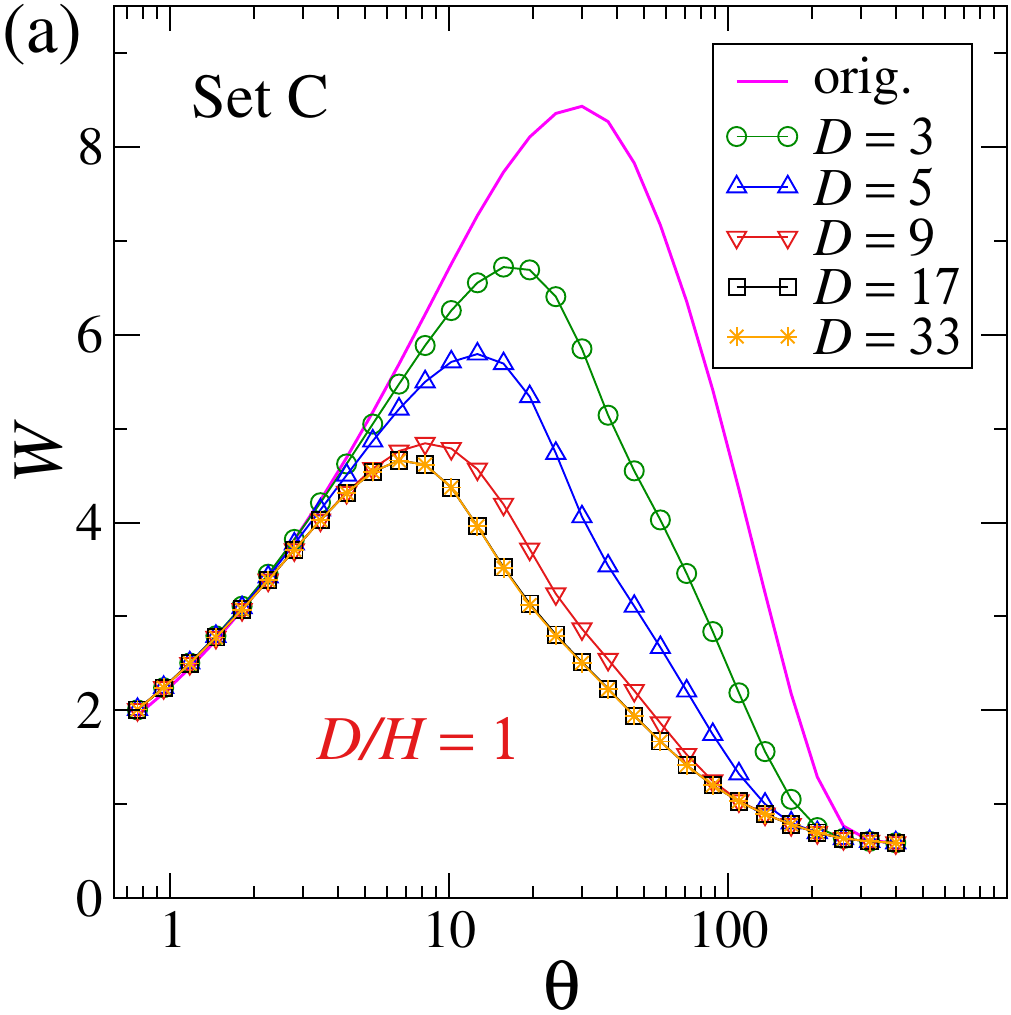}
\includegraphics[width=0.22\linewidth]{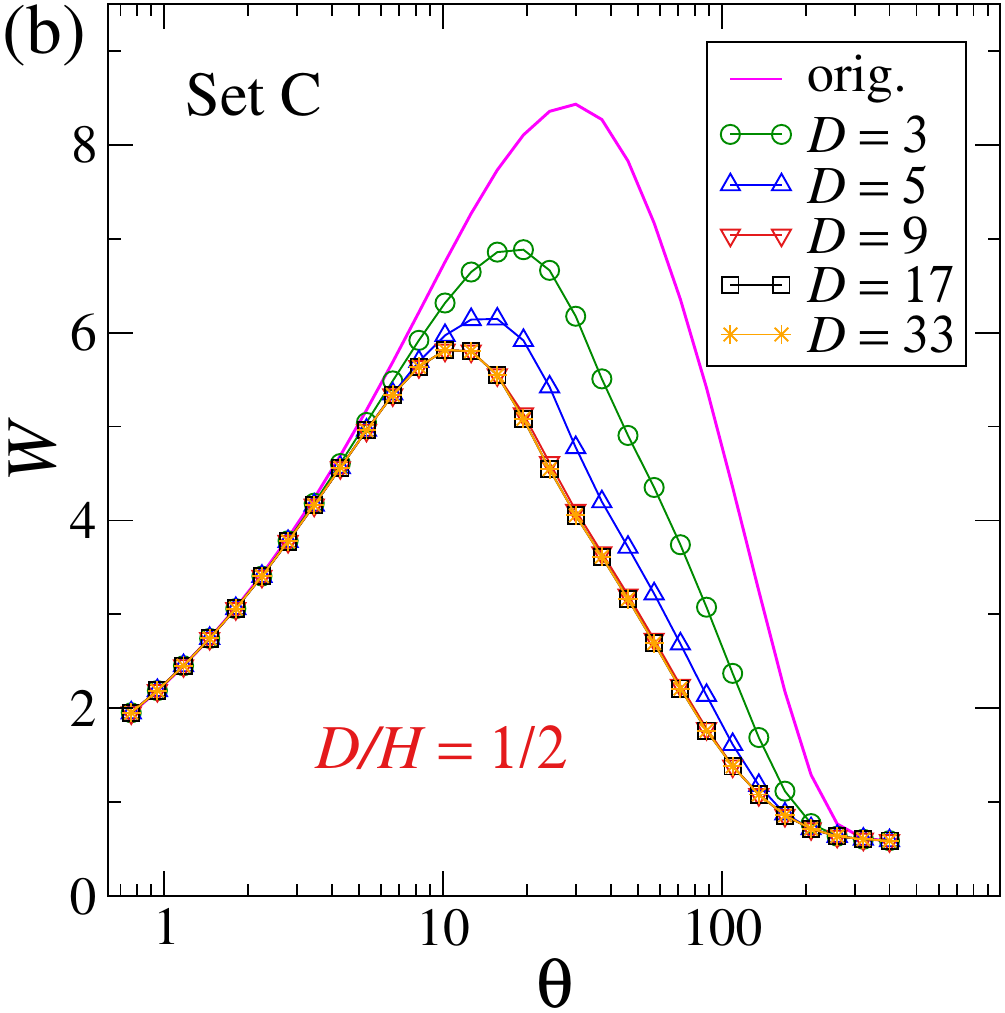}
\includegraphics[width=0.22\linewidth]{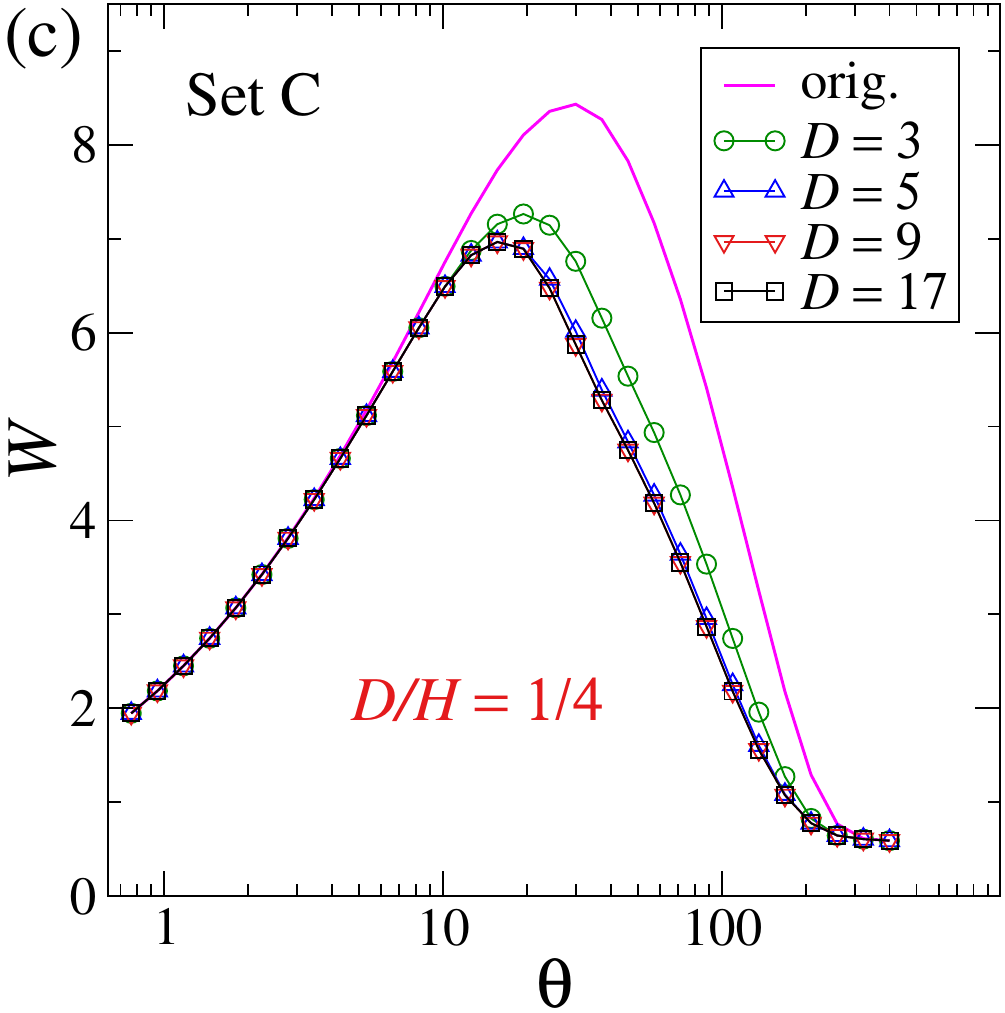}
\caption{Roughness as a function of the thickness for films grown with parameter set C (solid curve) and for transformed surfaces with model III and pyramidal probes (symbols).
The base widths are indicated by the legend.
The aspect ratios are: $\mathcal{R}=1$ (left), $1/2$ (middle) and $1/4$ (right panel).}
\label{roughnessIII}
\end{figure*}

Figures \ref{roughnessIII}(a)-(c) show the evolution of the roughness $W$ of surfaces grown with parameter set C and of the surfaces transformed with pyramidal probes with three aspect ratios $\mathcal{R}$ (model III).
In all cases, a shift of the maximal roughness to smaller roughness and a decrease of this maximal value is observed.
For a given aspect ratio, these changes increase with the size $D$: the wider the probe, the more difficult for it to penetrate the narrow gaps.
The changes also depend on the aspect ratio.
Using the less sharp probes, where $\mathcal{R}=1$, the smaller sizes lead to significant deviations in the maximal roughness, but for $D\geq17$ a single roughness curve is obtained with all tip sizes [Fig. \ref{roughnessIII}(a)].
Using the sharpest probes, where $\mathcal{R}=0.25$, the deviations from the original roughness curve are small and are nearly the same for all probes with $D\geq5$.
Similar results are obtained in films grown with the other parameter sets.

The above results can be qualitatively understood by inspection of Fig. \ref{probes}(c).
For a given aspect ratio, if the probe height is larger than the size of the surface fluctuations (i.e. $H\gtrsim W$), the probe collides with points of the surface that depend on the probe shape but do not depend on the probe size.
Thus, decreasing aspect ratios (sharpening the probe tips) are favorable for measuring the actual film surface.

\begin{figure}[!ht]
\includegraphics[width=0.24\linewidth]{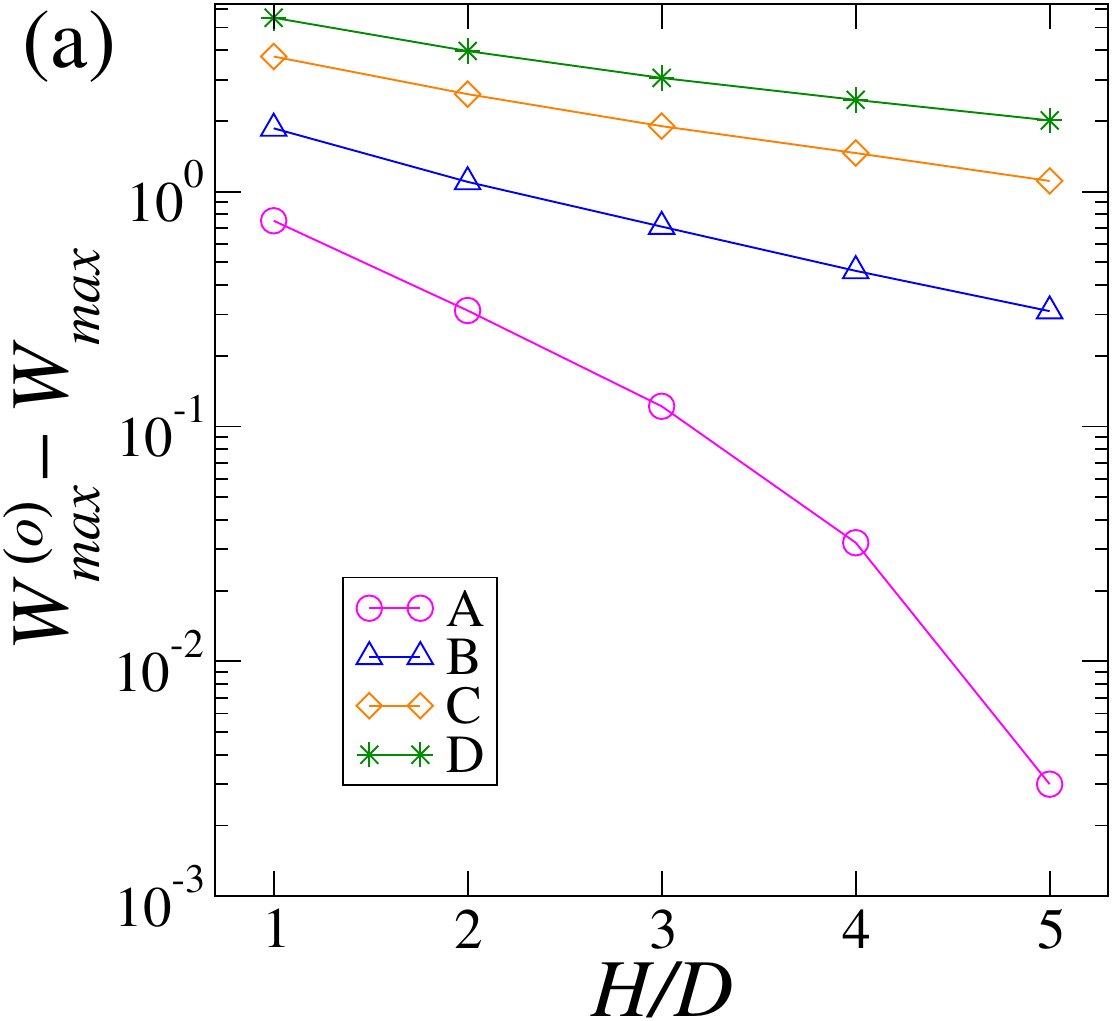}
\includegraphics[width=0.24\linewidth]{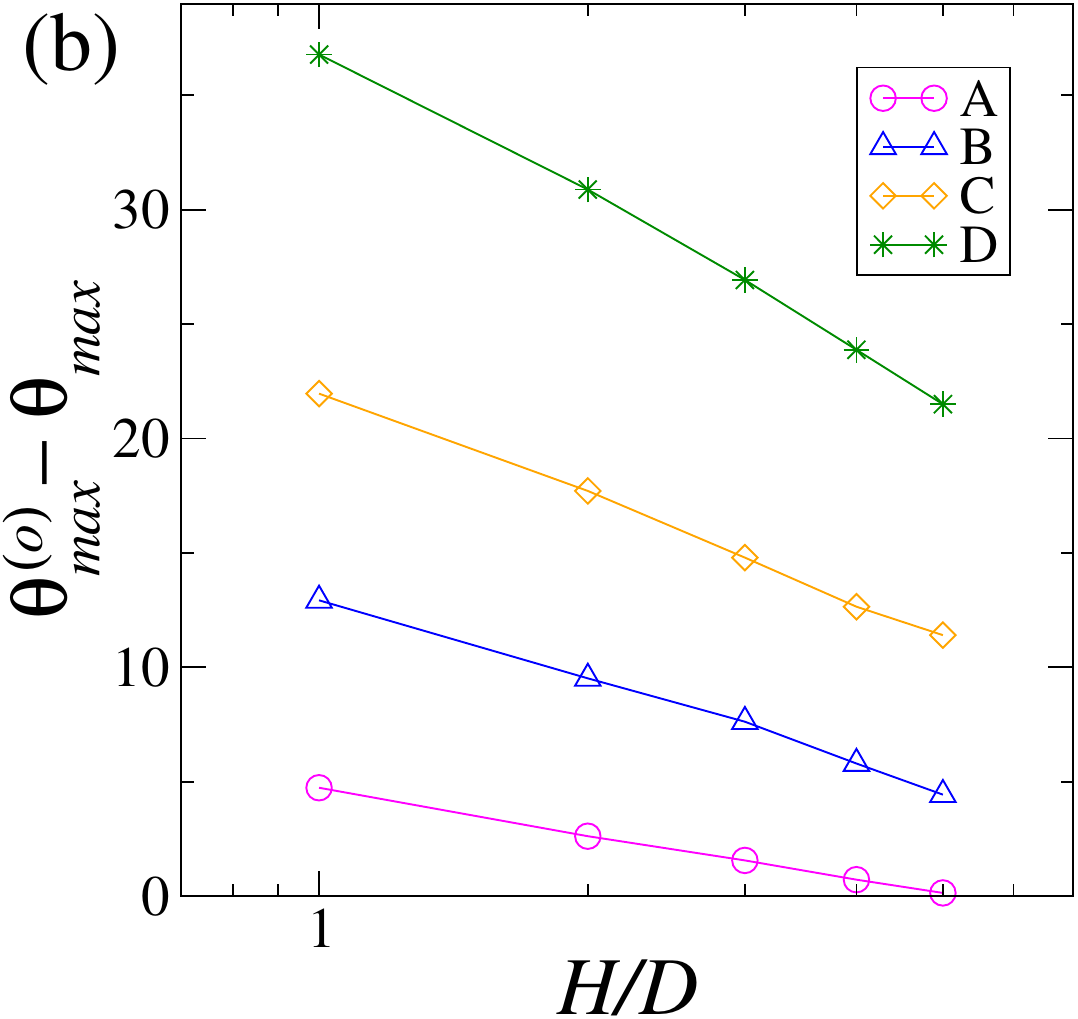}
\caption{(a) Difference between the maximal roughness of the original surfaces ($W_{max}^{(o)}$) and of the transformed ones ($W_{max}$) as a function of the reciprocal of the aspect ratio $H/D$ for model III and pyramidal probes.
(b) Difference between the corresponding coverages where the roughness is maximal ($\theta_{max}^{(o)}-\theta_{max}$).
The parameter sets are indicated.}
\label{maximoIII}
\end{figure}

Figure \ref{maximoIII}(a) shows the deviation of the maximal roughness as a function of the reciprocal of the aspect ratio of the probe for the four parameter sets; these deviations were obtained using the largest tips.
The deviation is roughly exponential in $H/D$ in sets B, C, and D; the same trend is not observed in set A, but in this case the deviation is always smaller than a lattice constant.
Figure \ref{maximoIII}(b) shows the corresponding deviations in the position of the maximal roughness, which has an approximately linear variation with $\log_{10}{(H/D)}$ in the studied range of $H/D$.

The results for conical tips are almost identical to those presented above for pyramidal tips of the same size and the same aspect ratio.
For these reasons, they are not presented here.

Although model III for the probe-surface interaction gives small deviations from the original surfaces in comparison with models I and II, it is important to observe that model III has assumptions that are difficult to be simultaneously implemented.
First, the width of the probe tip is smaller than the size of one deposited molecule, which is not common in SPM; instead, this assumption is more suitable for a scanning tunneling microscopy (STM) tip.
Second, the probe operates in contact mode, but an atomically thin tip may be damaged when it moves very close to a rough film surface (observe that STM does not operate with such contact).

\subsection{Insights from a geometric model}
\label{geometricmodel}

\begin{figure}[!ht]
\includegraphics[width=0.5\textwidth]{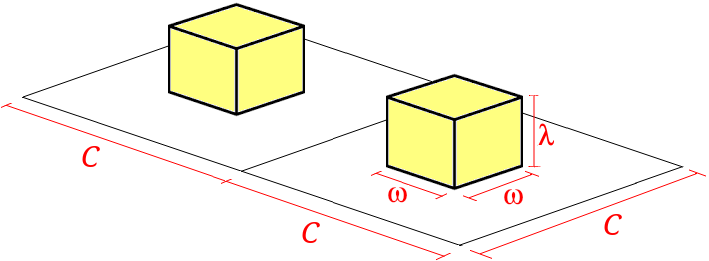}
\caption{Illustration of two islands in the geometric model.}
\label{figGeoModel}
\end{figure}

Here we consider a geometric model introduced in Ref. \protect\cite{graos}, in which islands are cuboids of lateral size $\omega$ and height $\lambda$ placed at the center of square capture zones of lateral size $C$; see Fig. \ref{figGeoModel}.
The parameters $\omega$, $\lambda$ and $C$ are dimensionless and must be interpreted as lengths given in units of the lattice parameter of a simple cubic geometry.
As shown in Ref. \citep{lapkinPRL2026}, the island growth and coalescence in the rough-to-smooth mode can be described by assuming that the width $\omega$ increases as a power law of the coverage $\theta$:
\begin{equation}
\omega=\omega_0 \theta^{\alpha} \quad ,\quad \alpha\in(0,1/2],
\label{omega}
\end{equation}
so that
\begin{equation}
\lambda=(C/\omega_0)^2\theta^{1-2\alpha} .
\label{lambda}
\end{equation}

Therefore, the fraction of covered substrate sites (height $\lambda$) is
\begin{equation}
f = {\left(\frac{\omega}{C}\right)}^2
\label{deff}
\end{equation}
and the height distribution $P(z)$ (i.e., the probability density function of finding a site with height $z$) is
\begin{equation}
P(z) = f \delta(z-\lambda) + (1-f)\delta(z) ,
\end{equation}
where $\delta$ denotes the Dirac delta function.
This yields a roughness
\begin{equation}
W^{(o)} = \lambda \sqrt{f-f^2},
\label{eqRugCuboid}
\end{equation}
which has a maximum at
\begin{equation}
 \theta_{max}^{(o)} = (1-\alpha)^{\frac{1}{2\alpha}} \left(\frac{C}{\omega_0} \right)^{\frac{1}{\alpha}}=(1-\alpha)^{\frac{1}{2\alpha}} \theta_{f}^{(o)}
\end{equation}
and which vanishes at a coverage
\begin{equation}
 \theta_{f}^{(o)} = \left(\frac{C}{\omega_0} \right)^{\frac{1}{\alpha}} ,
\end{equation} 
where
\begin{equation}
 W_{max}^{(o)} = (1-\alpha)^{\frac{1-\alpha}{2\alpha}} \sqrt{\alpha} \left(\frac{C}{\omega_0} \right)^{\frac{1}{\alpha}}.
\end{equation}
Of course, $W$ is also null at $\theta=0$ for a flat substrate.
As discussed in Ref. \citep{lapkinPRL2026}, from curves of $W \times \theta$ (for a given model or experimental system), these equations allows us to determine the values of $\alpha$ and $\omega_0/C$ in the geometric model.

Our main interest here is to analyze how these quantities for the original surfaces change after an SPM measurement, which is approximated by one of the transformation models described in the previous sections.
The height distribution and the corresponding roughness evolution can be modified by two effects:
\textit{i)} the \textit{apparent} fraction $f^{(a)}$ of covered substrate sizes will be larger than $f$, while the \textit{apparent} fraction, $f_\lambda^{(a)}$, of sites with height $\lambda$ can be larger or smaller than $f$ depending on the model for probe-surface interaction;
\textit{ii)} intermediate heights $0<z<\lambda$ may appear as a consequence of the tip width.
Letting $P_i(z)$ be the relative probability of finding an intermediate height $0<z<\lambda$, the generalized height distribution that accounts for all these effects reads
\begin{equation}
P(z) = f_{\lambda}^{(a)} \delta(z-\lambda) + \left(1 - f^{(a)} \right) \delta(z) + P_i(z) .
\label{eqHDgeral}
\end{equation}
From the normalization condition $\int_0^\lambda P(z)dz = 1$, one has $\int_0^\lambda  P_i(z) dz = f^{(a)}-f_{\lambda}^{(a)}$.

\subsubsection{Cylindrical tip - model II}
\label{cylindrical}

When the film is measured with a cylindrical tip of diameter $\ell$ and square cross-section, we still have the same scenario of the cuboidal islands above if $\ell$ is smaller than half of the width of the gap between the islands, i.e.
$\ell < (C-\omega)$.
However, the apparent island width increases to $\omega^{(a)}=\omega + 2\ell$ and the apparent fraction of covered sites increases to
\begin{equation}
f^{(a)}=f_{\lambda}^{(a)}=(\omega + 2\ell)^2/C^2 .
\label{facylindrical}
\end{equation}
There is no intermediate height between $0$ (gap) and $\lambda$ (island), so that $P_i=0$ in Eq. \ref{eqHDgeral}.
This means that the roughness is still given by Eq. \ref{eqRugCuboid}, but with $f$ replaced by $f^{(a)}$. After such a substitution, it is an easy task to show that $W$ now vanishes at
\begin{equation}
\theta_{f} = \theta_{f}^{(o)} \left[1-2\left(\frac{\ell}{C}\right) \right]^{\frac{1}{\alpha}}.
\label{eqThetafCyl}
\end{equation}
Therefore, the apparent range of coverages with an appreciable roughness decreases as $\ell$ augments.
This is indeed observed in the KMC results for cylindrical tips in Figs. \ref{roughnessII}(a)-(d), where `appreciable roughness' means $W > 1/2$. There is also a strong left-shift in the maximal roughness, whose location does decrease with $\ell$ as
\begin{equation}
\theta_{max} \simeq \theta_{max}^{(o)} \left[1-\frac{2-3\alpha+2\alpha^2}{(1-\alpha)^{\frac{3}{2}}}\left(\frac{\ell}{C}\right) + \ldots \right]^{\frac{1}{\alpha}}.
\label{eqThetamaxCyl}
\end{equation}
Inserting this result into Eq. \ref{eqRugCuboid} and expanding it up to first order in $\ell$, one gets
\begin{equation}
W_{max} \simeq W_{max}^{(o)} \left[ 1-\frac{2-4\alpha}{\sqrt{1-\alpha}}\left(\frac{\ell}{C}\right) + \ldots \right]^{\frac{1}{\alpha}},
\label{eqRugmaxCyl}
\end{equation}
which is also consistent with the decreasing maxima shown in the KMC simulations of Figs. \ref{roughnessII}(a)-(d).

It is also important to notice that, for small $\theta$, the roughness behaves as
\begin{equation}
W \simeq W^{(o)} \left\{1+2\theta^{-\alpha}\left[1-\left(\frac{\omega_0}{C}\right)^2 \theta^{2\alpha}\right]\left(\frac{\ell}{\omega_0}\right) \right\},
\end{equation}
where $W^{(o)}$ is width of the original film (given by Eq. \ref{eqRugCuboid}).
Thus, $W$ increases with $\ell$ in this low coverage regime, in agreement with the simulations in Figs. \ref{roughnessII}(a)-(d).

\begin{figure*}[!ht]
	\includegraphics[width=0.25\linewidth]{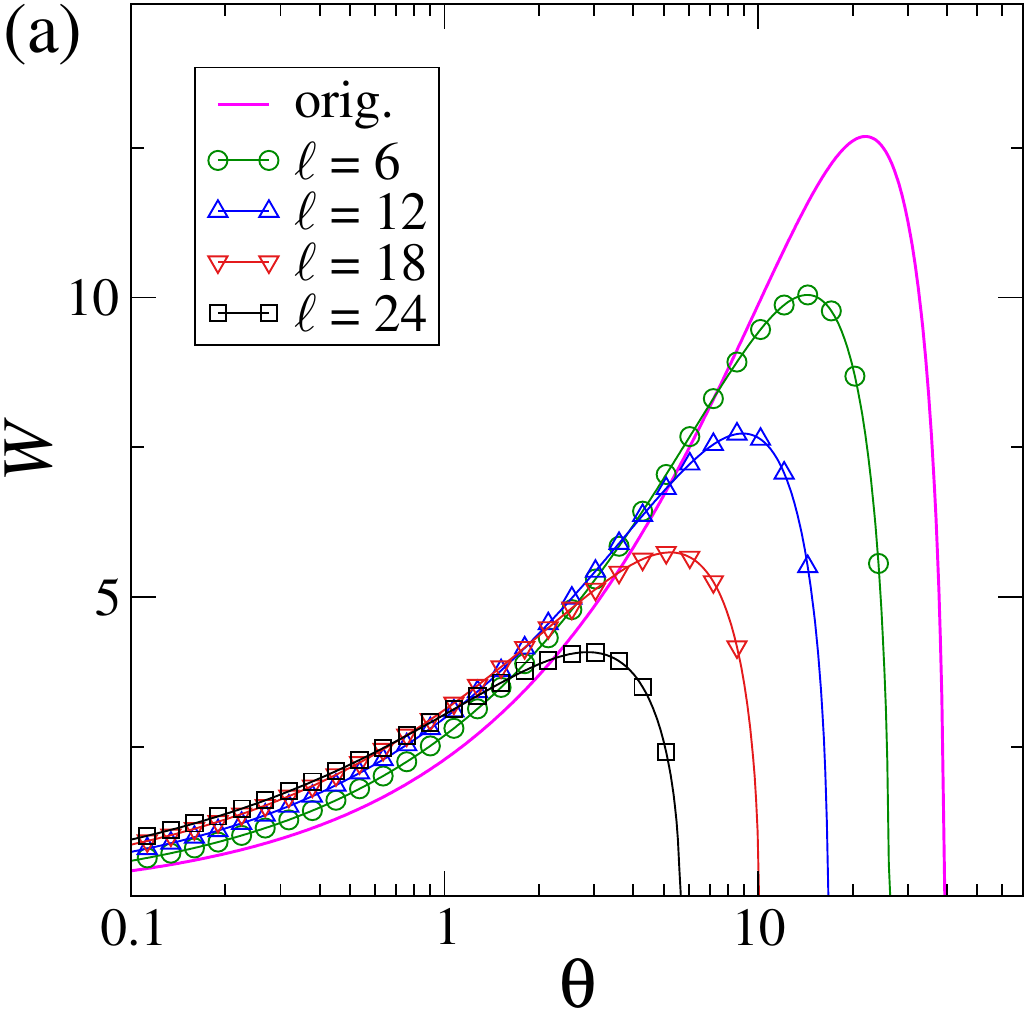}
	\includegraphics[width=0.25\linewidth]{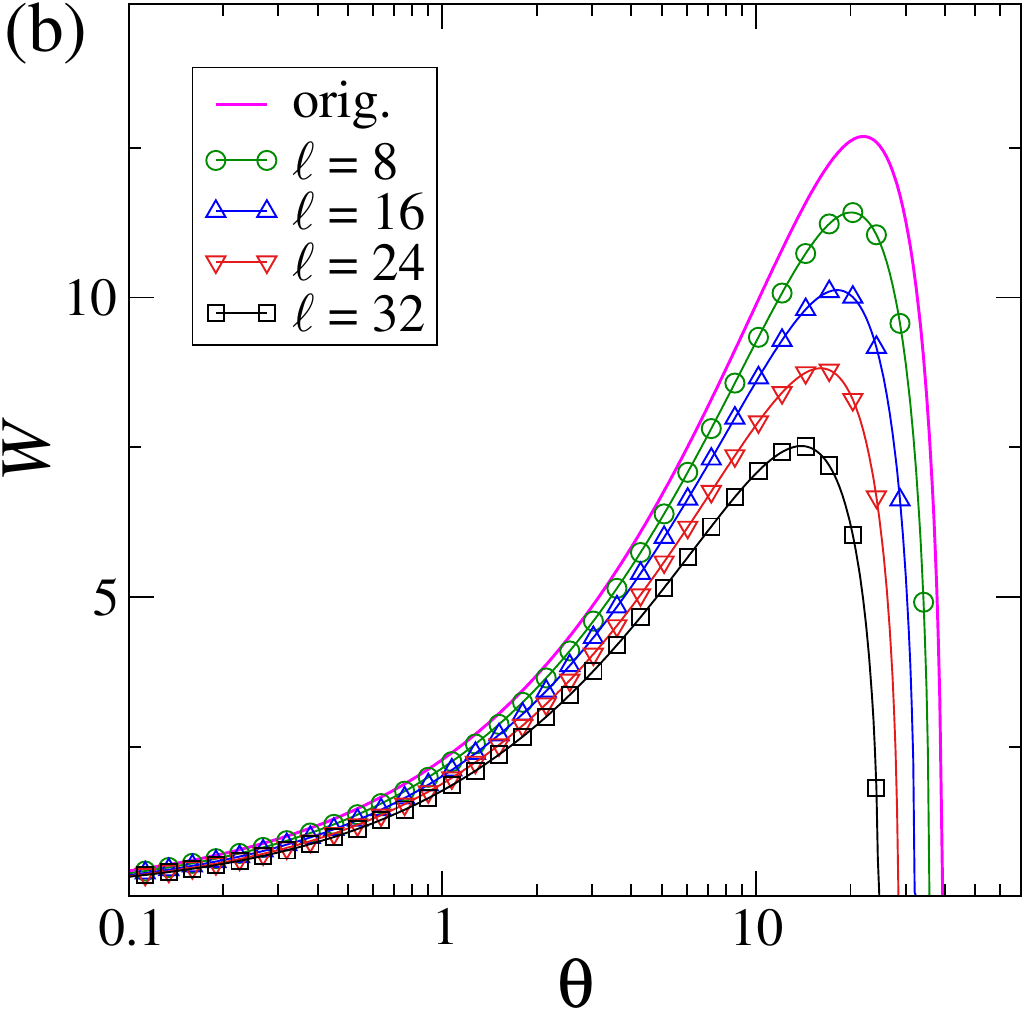}
	\includegraphics[width=0.25\linewidth]{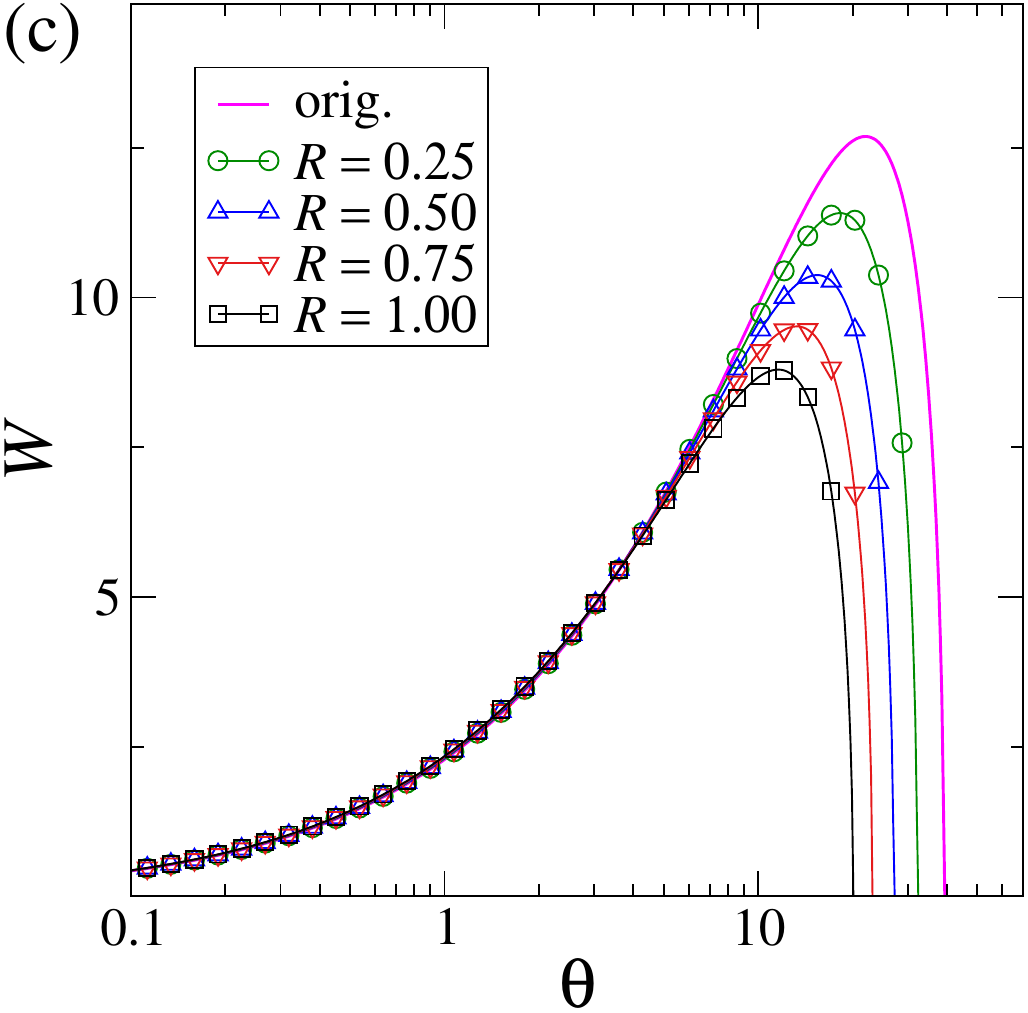}
	\caption{Roughness as a function of thickness for the geometrical model as measured by probes with different shapes. The solid magenta curve represents the original geometrical model, obtained with $\omega_0=50$, $C=150$, and $\alpha=1/4$. Panels (a), (b), and (c) show, respectively, the effect of the cylindrical, mean-height, and pyramidal probe approaches.}
	\label{curvasnumericas}
\end{figure*}

These changes may also be appreciated by calculating numerically the roughness variation with the thickness for a given set of parameters of the geometrical model.
Here we use $\omega_0=50$, $C=150$, and $\alpha=1/4$, with the evolutions of island width and height given by Eqs. (\ref{omega}) and (\ref{lambda}), and the roughness given by Eq. (\ref{eqRugCuboid}) with the fraction of covered sites of Eq. (\ref{facylindrical}).
Figure \ref{curvasnumericas}(a) shows that increasing $\ell$ leads to a decrease in the maximal roughness and shifts the curves to the left, as observed in the KMC simulations of Figs. \ref{roughnessII}(a)-(d).

An important difference between the geometrical model and the KMC simulations is the much faster decay of the roughness after the peak in the former; compare, for instance, the original curves in Figs. \ref{comparison} and \ref{curvasnumericas}(a).
An improvement of the geometrical model, as suggested in Ref. \protect\cite{lapkinPRL2026}, may lead to better agreement with the KMC data in that regime, but with weak effects on the peak height and on the peak position.

\subsubsection{Mean height approach - model I}

Here we consider narrow probe tips with square cross-sections and sides in the same directions of the islands.
The probe sizes are assumed to be smaller than the island widths, i.e. $\ell < \omega$.

The apparent surface fraction with the island height $\lambda$ decreases with $\ell$ as
\begin{equation}
f_{\lambda}^{(a)} = (\omega-\ell)^2/C^2 ,
\label{flambdamean}
\end{equation}
contrary to the increase obtained in model II.
The apparent fraction of covered sites increases to
\begin{equation}
f^{(a)} = (\omega+\ell)^2/C^2 .
\label{famean}
\end{equation}
Moreover, apparent intermediate heights forming lateral facets of slope $(\lambda/\ell)$ will appear as an effect of the tip, leading the measured islands to have approximate pyramidal shapes (except at the corners) truncated at height $\lambda$.
Hence, we will proceed using such a truncated pyramid as an approximation to the measured island.

For a square frustum of height $\lambda$, base side $b$ and top side $a\leq b$, placed on a square substrate of side $C>b$, the probability of finding a height $0<z<\lambda$ is:
\begin{equation}
P_i(z)=\frac{2(b-a)}{\lambda C^2}\left(b-\frac{b-a}{\lambda}z\right) = \frac{4\ell}{\lambda C^2}\left(\omega+\ell-\frac{2\ell}{\lambda}z\right) ,
\label{eqProbInt}
\end{equation}
where the right hand side is obtained with $b=(\omega+\ell)$ and $a=(\omega - \ell)$ in the approximation of the measured island.
By replacing this into Eq. (\ref{eqHDgeral}) and calculating the first moments of $P(z)$, we arrive at the roughness
\begin{equation}
W = h \sqrt{\frac{f^{(a)}+2S+3 f_{\lambda}^{(a)}}{6}-\frac{\left(f^{(a)}+S+f_{\lambda}^{(a)}\right)^2}{9}},
\label{eqRugFrustum}
\end{equation}
where $S\equiv \sqrt{f^{(a)}f_{\lambda}^{(a)}}$.
Note that this reduces to Eq. (\ref{eqRugCuboid}) if $f^{(a)}=f_{\lambda}^{(a)}=f$.

It is not possible to find a simple expression for the coverage $\theta_f$ at which $W=0$.
However, the expansion of Eqs. (\ref{flambdamean}) and (\ref{famean}) for small $\ell/C$ can be inserted into Eq. \ref{eqRugFrustum} to give
\begin{equation}
\theta_{f} = \theta_{f}^{(o)} \left[1-\frac{1}{3}\left(\frac{\ell}{C}\right) -\frac{1}{3}\left(\frac{\ell}{C}\right)^2 - \cdots \right]^{\frac{1}{\alpha}}.
\label{thetafmean}
\end{equation}
Using the same expansion and writing down the roughness of Eq. (\ref{eqRugFrustum}) in terms of $\theta$, the maximal roughness is 
\begin{equation}
W_{max} \simeq W_{max}^{(o)} \left[ 1-\frac{1}{3\sqrt{1-\alpha}}\left(\frac{\ell}{C}\right) - \ldots \right]^{\frac{1}{\alpha}}
\label{Wmaxmean}
\end{equation}
and is obtained at coverage
\begin{equation}
\theta_{max} \simeq \theta_{max}^{(o)} \left[1-\frac{2-3\alpha}{6(1-\alpha)^{\frac{3}{2}}}\left(\frac{\ell}{C}\right) - \ldots \right]^{\frac{1}{\alpha}} .
\label{thetamaxmean}
\end{equation}

The roughness peak and the position of this peak decrease with $\ell$ as observed in the KMC simulations of Figs. \ref{roughnessI}(a)-(d).
It is noteworthy that the prefactors of the linear term $(\ell/C)$ found here for $\theta_{f}$ and $\theta_{max}$ [Eqs. (\ref{thetafmean}) and (\ref{thetamaxmean})] are much smaller (by factors $\geq 6$) than those found just above for model II [Eqs. (\ref{eqThetafCyl}) and (\ref{eqThetamaxCyl})].
This explains why much stronger shifts of the roughness peaks are observed in the transformations with model II (cylindrical tips), in agreement with the KMC simulations [Figs. \ref{roughnessI}(a)-(d) and \ref{roughnessII}(a)-(d)].
However, the prefactor found here for $W_{max}$ [Eq. (\ref{Wmaxmean})] may be larger or smaller than the one for model II [Eq. (\ref{eqRugmaxCyl})] depending on whether $\alpha$ is larger or smaller than $5/12$.
Indeed, in the simulations, significant decreases of $W_{max}$ with $\ell$ are observed with both transformation models.

Regarding the roughness behavior for small $\theta$, we obtain
\begin{equation}
W \simeq W^{(o)} \left\{1-\frac{\theta^{-\alpha}}{3}\left[1+\left(\frac{\omega_0}{C}\right)^2 \theta^{2\alpha}\right]\left(\frac{\ell}{\omega_0}\right) \right\}.
\end{equation}
Although this equation looks similar to that for the cylindrical tips, now $W$ decreases with $\ell$, in agreement with the KMC simulations of Figs. \ref{roughnessI}(a)-(d).

Figure \ref{curvasnumericas}(b) illustrates how the probe size affects the whole behavior of the roughness given by Eq. \ref{eqRugFrustum} for the geometrical model with $\omega_0=50$, $C=150$, and $\alpha=1/4$.
The decrease of the roughness peak $W_{max}$ is remarkable, whereas the leftward shift of the peak position $\theta_{max}$ is only marginal, consistently with the preceding discussion and with the KMC simulations of Figs. \ref{roughnessI}(a)-(d).

\subsubsection{Pyramidal and conical tips - model III}
\label{pyramidal}

Here we consider pyramidal probes whose bases have width $\ell$ and aspect ratio $\mathcal{R}=\ell/H$ [Fig. \ref{probes}(c)].
The apparent islands measured with these probes are square frustums, so that the roughness is again given by Eq. (\ref{eqRugFrustum}) derived for model I.
However, now the apparent fraction of covered sites (height $>0$) is
\begin{equation}
f^{(a)} = (\omega +\lambda \mathcal{R})^2/C^2 ,
\label{fapyramid}
\end{equation}
while the fraction of sites with the maximal height $\lambda$ is the same as in the original surface, $f_{\lambda}^{(a)} = f= \omega^2/C^2$ [Eq. (\ref{deff})].
As in the KMC simulations, we are considering that the sides of the pyramidal tip are parallel to those of the islands.


For conical probes, the measured islands are also similar to truncated pyramids, but with rounded corners.
For both probe shapes, lateral facets of slope $2/\mathcal{R}$ appear as a feature introduced by the finite probe sizes.
From this observation, the roughness can be calculated for both shapes.

Here we explicitly calculate the roughness for the conical case.
Each of the four facets of the measured truncated pyramids have width $\omega$ and each corner is a quarter of cone with slope $2/\mathcal{R}$ and height $\lambda$, so that the base has a radius $\lambda\mathcal{R}/2$.
This means that the fraction of covered sites is
\begin{equation}
f^{(a)}=[\omega^2 + 2\omega \lambda \mathcal{R} + \pi \lambda^2\mathcal{R}^2/4]/C^2 ,
\label{facone}
\end{equation}
while the fraction of sites with the maximal height $\lambda$ remains the same, $f_{\lambda}^{(a)}=f=\omega^2/C^2$.
The probability of intermediate heights is
\begin{equation}
P_i(z) = \frac{1}{C^2}\left[2\omega \mathcal{R} + \frac{\pi \mathcal{R}^2}{2}(\lambda-z)\right],
\end{equation}
which leads to the roughness
\begin{equation}
W = \lambda \sqrt{\frac{f^{(a)}+5f_{\lambda}^{(a)}+A}{6}-\left(\frac{2f^{(a)}+4f_{\lambda}^{(a)}+A}{6}\right)^2},
\label{eqRugRoundedPyram}
\end{equation}
where
\begin{equation}
A = \frac{8f_{\lambda}^{(a)}}{\pi}\left(\sqrt{1+\frac{\pi\left(f^{(a)}-f_{\lambda}^{(a)}\right)}{4f_{\lambda}^{(a)}}}-1\right).
\end{equation}
Once again, Eq. \ref{eqRugCuboid} is recovered if $f^{(a)}=f_{\lambda}^{(a)}=f$.
This roughness vanishes at
\begin{equation}
\theta_{f} = \theta_{f}^{(o)} \left[1-\frac{2}{3}\left(\frac{\ell \lambda_f^{(o)}}{C H}\right) + \frac{12-17\alpha}{18 \alpha}\left(\frac{\ell \lambda_f^{(o)} }{C H}\right)^2 + \cdots \right]^{\frac{1}{\alpha}},
\label{thetafcone}
\end{equation}
where $\lambda_f^{(o)} = (C/\omega_0)^2 \left(\theta_{f}^{(o)}\right)^{1-2\alpha}=\theta_{f}^{(o)}$.

A very similar expression is obtained for pyramidal tips, but with the second-order coefficient changed to $(12-21.5\alpha)/(18\alpha)$.
In general, the results for tips with conical and pyramidal shapes will be the same up to first-order in $\mathcal{R}$.

Maximizing the roughness with respect to the coverage $\theta$, we find
\begin{equation}
\theta_{max} \simeq \theta_{max}^{(o)} \left[1-\frac{6-13\alpha+9\alpha^2}{6(1-\alpha)^{\frac{3}{2}}}\left(\frac{\ell \lambda_{max}^{(o)}}{C H}\right) - \ldots \right]^{\frac{1}{\alpha}} ,
\end{equation}
where $\lambda_{max}^{(o)} = (C/\omega_0)^2 \left(\theta_{max}^{(o)}\right)^{1-2\alpha}$.
At this maximum, the roughness is approximately
\begin{equation}
W_{max} \simeq W_{max}^{(o)} \left[ 1-\frac{2-3\alpha}{3\sqrt{1-\alpha}}\left(\frac{\ell \lambda_{max}^{(o)}}{C H}\right) - \ldots \right]^{\frac{1}{\alpha}} .
\label{Wmaxcone}
\end{equation}

In Eq. (\ref{thetafcone}), the coefficient of the first-order correction term in $\ell/C$ is twice as large than in model I (mean height case), but smaller by a factor $3$ in comparison with model II (maximal height case for cylindrical probes).
Moreover, it competes with the positive second-order term, which may dominate for small $\alpha$.
This is consistent with the KMC simulations of Figs. \ref{roughnessIII}(a)-(c), in which the roughness becomes small ($W\approx 1/2$, which is close to LBL growth) at nearly the same $\theta$ for all aspect ratios.
The prefactor of the linear correction in $\theta_{max}$ found here is also smaller than that for model II (cylindrical probes), but larger than that of model I (the mean height method).
This also agrees with the intermediate shifts observed in the KMC results for pyramidal probes [Figs. \ref{roughnessIII}(a)-(c)] or for conical probes.
In Eq. (\ref{Wmaxcone}), the correction coefficient of $W_{max}$ for $1/3 < \alpha < 4/9$ is the smallest one among all transformation models; otherwise, it is intermediate between models I and II.

For small $\theta$, the roughness is now given by
\begin{equation}
W \simeq W^{(o)} \left\{ 1 + \left[\frac{1}{3}\left(\frac{C}{\omega_0} \right)^2 \theta^{1-3\alpha} - \frac{2}{3}\theta^{1-\alpha}\right]\left(\frac{\mathcal{R}}{\omega_0}\right) + \ldots \right\}.
\end{equation}
Here the leading correction is proportional to $\sim \theta^{1-3\alpha}$.
For $\alpha<1/3$, this correction is negligible for small $\theta$, indicating that $W$ will be independent of the aspect ratio $\mathcal{R}$ in this regime.
This agrees with the simulations shown in Figs. \ref{roughnessIII}(a)-(c), in contrast with the other transformation models.

Fig. \ref{curvasnumericas}(c) shows the roughness evolution obtained with the transformation model III (pyramidal probes) applied to the geometrical model with $\omega_0=50$, $C=150$, and $\alpha=1/4$; similar result is obtained with the conical probes.
The leftward shift is more pronounced than that obtained with model I (mean height approach), but smaller than the shift produced by model II (cylindrical probe), in agreement with the above discussion and in parallel with the KMC simulations.

\subsection{Application to $\text{C}_{60}$ films}
\label{application}

Using XRR to analyze the surface of evaporated $\text{C}_{60}$ films, Lapkin \textit{et al.} showed that they follow the rough-to-smooth growth mode with a roughness peak of $\approx3$~nm at a thickness of $\approx5$~nm \citep{lapkinPRL2026}.
The uncertainty in the roughness measured with this technique is generally smaller than $0.1$~nm.
AFM images of the same films were performed in tapping mode.
They showed a sharp roughness peak of $\approx1.6$~nm at a thickness of $\approx3$~nm, which is followed by a roughness plateau of $\approx1.3$~nm that extends until $\approx9$~nm of thickness; see Fig. S4 of the Supplemental Material of Ref. \protect\cite{lapkinPRL2026}.

Considering the good accuracy of the XRR measurements, we understand that they can be viewed as the most accurate representation of the actual fluctuations of the film surfaces.
The roughness peak obtained in the AFM study was lower and was located at a smaller thickness than the XRR peak.
This is qualitatively consistent with the predictions of our models (I, II, and III), which encompass different probe-surface interactions.

Results for a complementary $\text{C}_{60}$ film were presented in Fig. S7 of the Supplemental Material of Ref. \protect\cite{lapkinPRL2026}.
Using XRR, a roughness peak of $\approx2.8$~nm was obtained in a broad range of thicknesses, from $\approx3$~nm to $\approx6$~nm.
With AFM, the roughness peak of $\approx1$~nm is displaced to thicknesses from $\approx8$~nm to $\approx9$~nm, i.e., larger thicknesses than those of the XRR peak.

The interpretation of this AFM result in the light of our models suggests that the actually probed films have a roughness peak above $\approx1$~nm and that this peak is located at a thickness of $9$~nm or larger.
This thickness is much larger than the positions of the roughness peak obtained with XRR.
This suggests that island coarsening and possible dewetting occurred between the XRR and the AFM measurement, so that the islands probed with AFM had larger widths and their coalescence was delayed to larger thicknesses.
Thus, our models may also help to infer the existence of post-deposition dynamics.

\section{Conclusion}
\label{Conc}

When thin solid films are deposited, a frequently observed feature is continuous surface roughening, which may be explained in terms of theories of kinetic roughening or mound coarsening.
Microscopy and spectroscopy techniques have been used in these studies, with SPM playing an important role in showing real-space fluctuations at several length scales.
In order to enhance the accuracy of SPM, some previous works analyzed possible deviations in the measured roughness due to interactions between probes of finite widths and the scanned surfaces which show this type of roughening.
However, in the last two decades, there appeared evidence that some inorganic and organic films follow a rough-to-smooth growth mode in which the surface roughness does not increase monotonically.
Instead, multilayer islands are formed on weakly interacting substrates and, after the islands coalesce, a continuous film with low roughness appears.
This growth mode was quantitatively explained in ultrathin organic films by a combination of a geometric model and KMC simulations.
Here, we studied the evolution of the roughness measured by SPM using probes of finite size that interact with film surfaces showing that growth mode.
KMC simulations were used to generate those films and different models of probe-surface interactions were used to produce the SPM-transformed surfaces.
The obtained results were theoretically explained in the context of the geometric model applied to that growth mode.
This study is important due to the increased technological interest in films with nanoscale thickness and the need for accurate characterization of the structure of these materials.

Three models of probe-surface interaction were considered to generate transformed surfaces from the original film surfaces.
For modeling the non-contact mode (model I), the transformed profile at each point was obtained by averaging surface heights in a region of the same width as the probe tip; this method is supported by a recent study of electrostatic force microscopy \citep{amorim2023}.
For modeling contact modes, cylindrical probes with hard-core interactions with the film surfaces provide transformed surfaces in which the height at each position is the maximal value inside a box with the shape of the cylinder cross-section (model II).
Additionally, the contact mode is considered with pyramidal and conical probes whose tips are oriented oppositely to the growth direction and which also have hard-core interactions with the film surfaces (model III).

We first performed KMC simulations to generate the evolution of film surfaces following the rough-to-smooth mode.
Using model I, the transformed surfaces also show that growth mode if the probe width $\ell$ is smaller than the average island distance, with roughness peaks decreasing with $\ell$ and small displacements of that peak to smaller thicknesses.
In the same condition for the probe width, the surfaces transformed with model II (cylindrical probes in contact mode) also show the rough-to-smooth mode with roughness peaks decreasing with $\ell$, but there are significant displacements of these peaks to smaller thicknesses.
Thus, SPM images obtained with probes that approximately follow these models may lead to subestimation of the non-uniformity of the film thickness in deposits with coalescing islands, which may affect their expected physical properties.
The situation is worse when the probe width is of the same order of magnitude as the average island distance because the transformation of the surface with models I and II does not lead to a roughness peak and possibly yields a decreasing roughness with model II.

Finally, with model III for pyramidal and conical probes in contact mode, the above deviations in the roughness peak and in the position of this peak are observed up to a certain width $\ell$ that depends on the aspect ratio of the probes.
For sharp tips (small ratios between base width and probe height), the deviations are observed only for small $\ell$.
However, the conditions of model III are hard to implement because the probe tips are atomically thin and, when operating in contact mode, they may be damaged when scanning the film surface.

The geometric model that approximately represents the rough-to-smooth mode assumes power-law variation of island widths and heights before island coalescence.
When the surfaces generated with this geometric model are transformed by models I, II, and III, the resulting roughness evolution shows the same qualitative trends observed in the KMC simulations: the height of the roughness peak and the thickness where it appears decrease as the tip width increases.
Moreover, the geometric model also explains the much smaller deviations of the peak position in model I when compared to model II.

In a recent study of $\text{C}_{60}$ films showing the rough-to-smooth growth mode, the roughness peaks obtained with AFM were significantly lower than those obtained with XRR.
Assuming that the non-invasive character of XRR allows a more accurate measurement of the film surface, the observed change in the AFM roughness peaks is consistent with our models.
However, in one of the studied samples, the thickness of the AFM roughness peak was larger than the thickness of the XRR peak.
In light of our models, we expect that island coarsening and possible dewetting occurred in this sample in the time interval between the XRR and the AFM studies. 
Thus, our results may also help to detect effects of post-deposition dynamics.

\section*{Supplementary Material}

The Supplementary Material contains additional images of original and transformed surfaces.

\section*{Acknowledgements}
FDAAR acknowledges support from the Brazilian agencies CNPq (305570/2022-6), FAPERJ (E-26/200.282/2026, E-26/210.062/2023), and CAPES (88881.700849/2022-01).
TJO acknowledges support from CNPq and FAPEMIG.
ISSC acknowledges support from CNPq (408929/2025-1, 305367/2026-9).

\section*{Data Availability}

The data that support the findings of this study are available from the corresponding author upon reasonable request.

\section*{Conflict of interest}

The authors declare that they have no known competing financial interests or personal relationships that could have appeared to influence the work reported in this paper.

\section*{Author Contributions}

\par\noindent Ismael S.S. Carrasco: Writing - review \& editing, Software, Visualization, Validation.

\par\noindent Tiago J. Oliveira: Writing - original draft, Methodology, Visualization, Investigation, Formal analysis, Data curation.

\par\noindent F\'abio D. A. Aar\~ao Reis: Writing - original draft, Writing - review and editing, Conceptualization, Methodology, Investigation, Formal analysis, Resources, Project administration, Funding acquisition.

\bibliography{interfaces}

\clearpage

\appendix

\setcounter{figure}{0}
\renewcommand{\thefigure}{S\arabic{figure}}

\section{Additional snapshots of original and transformed surface profiles}
\label{additional}

\begin{figure}[!ht]
	\includegraphics[width=0.9\textwidth]{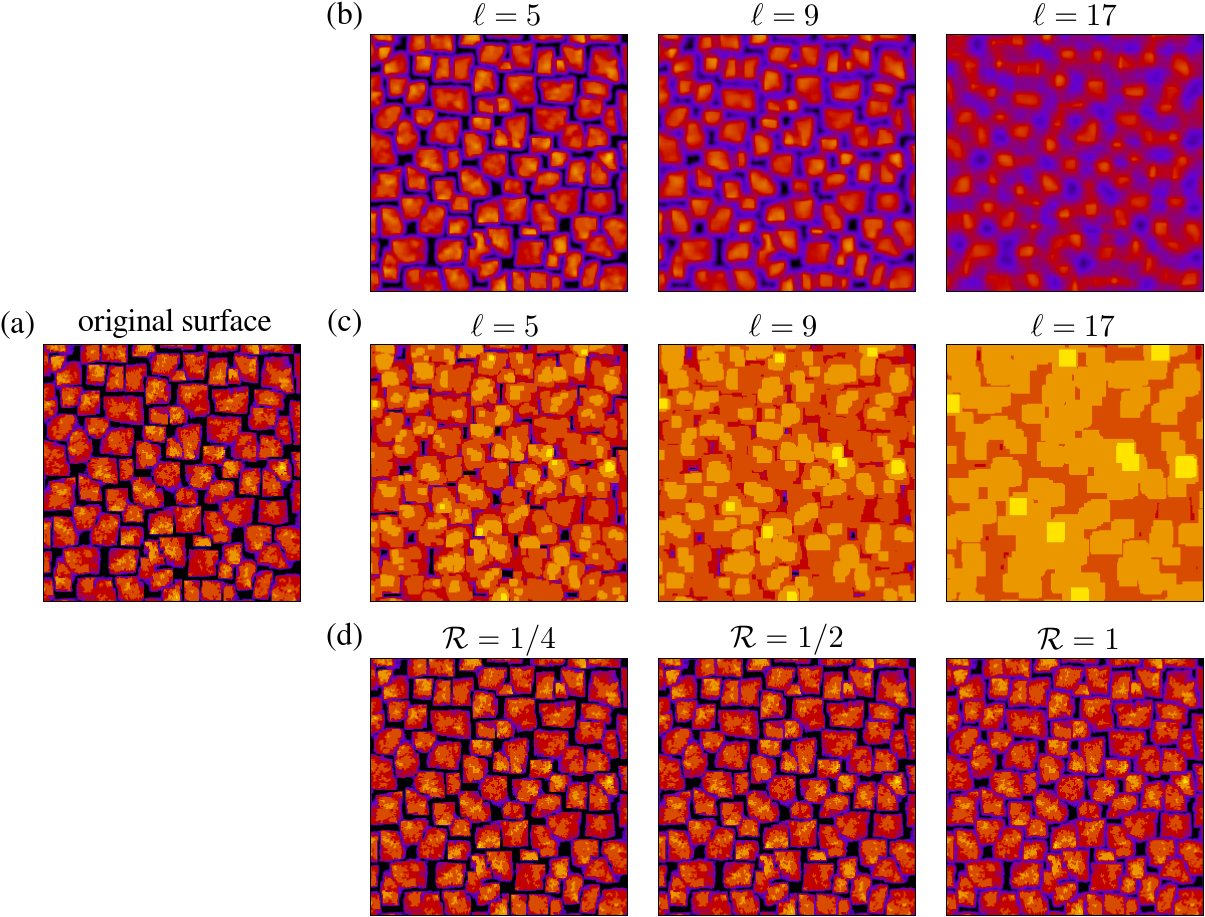}
	\caption{Snapshots of size $256\times256$ taken from: (a) the original surface grown with set A at coverage $\theta = 4.7287$; (b), (c), and (d) the transformed surfaces obtained from (a) using models I, II, and III, respectively.
		The results in (b) and (c) are for square boxes with the indicated sizes.
		Case (d) is for pyramidal tips with base $D=9$ and the indicated aspect ratios.}
	\label{snapshotsA}
\end{figure}

\begin{figure}[!ht]
	\includegraphics[width=0.9\textwidth]{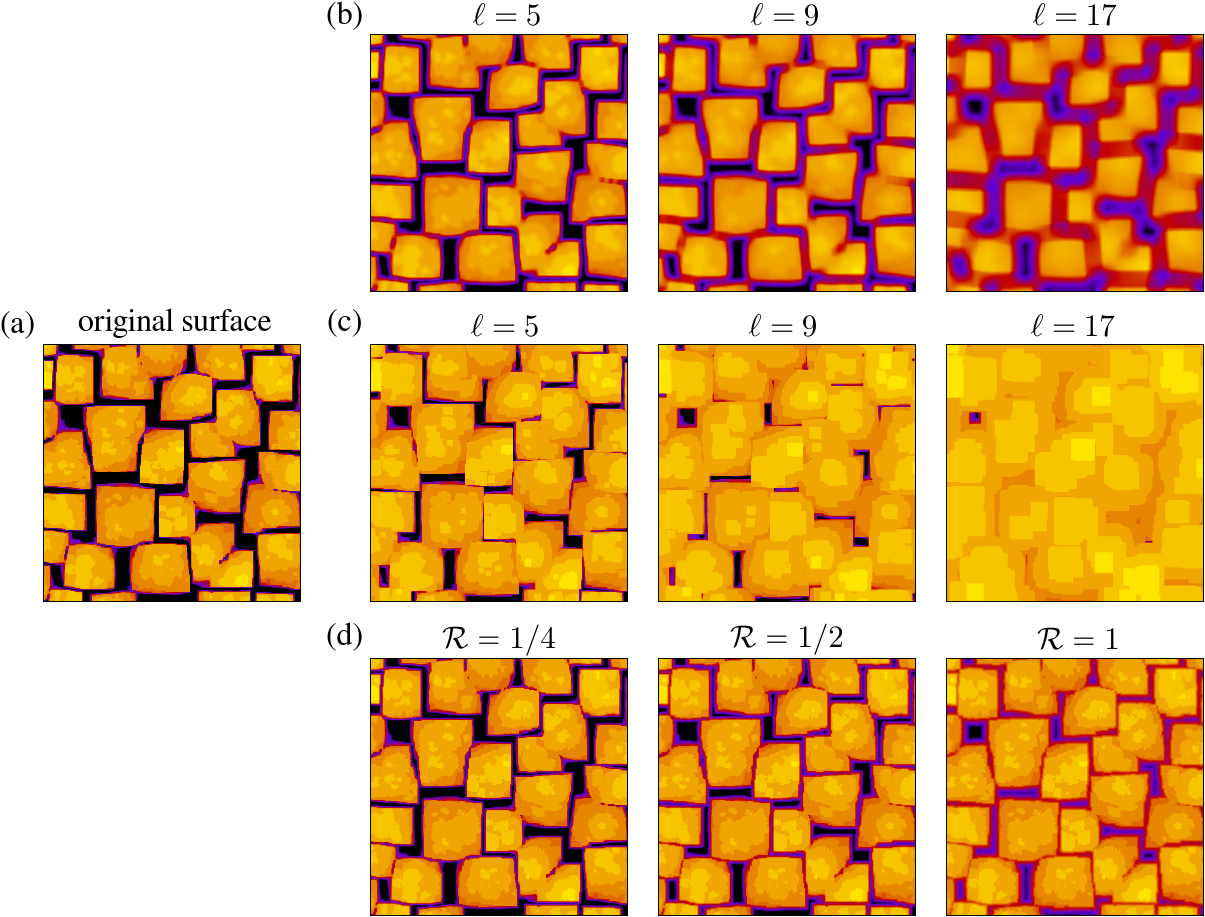}
	\caption{Snapshots of size $256\times256$ taken from: (a) the original surface grown with set C at coverage $\theta = 15.6995$; (b), (c), and (d) the transformed surfaces obtained from (a) using models I, II, and III, respectively.
		The results in (b) and (c) are for square boxes with the indicated sizes.
		Case (d) is for pyramidal tips with base $D=22$ and the indicated aspect ratios.}
	\label{snapshotsC}
\end{figure}

\begin{figure}[!ht]
	\includegraphics[width=0.9\textwidth]{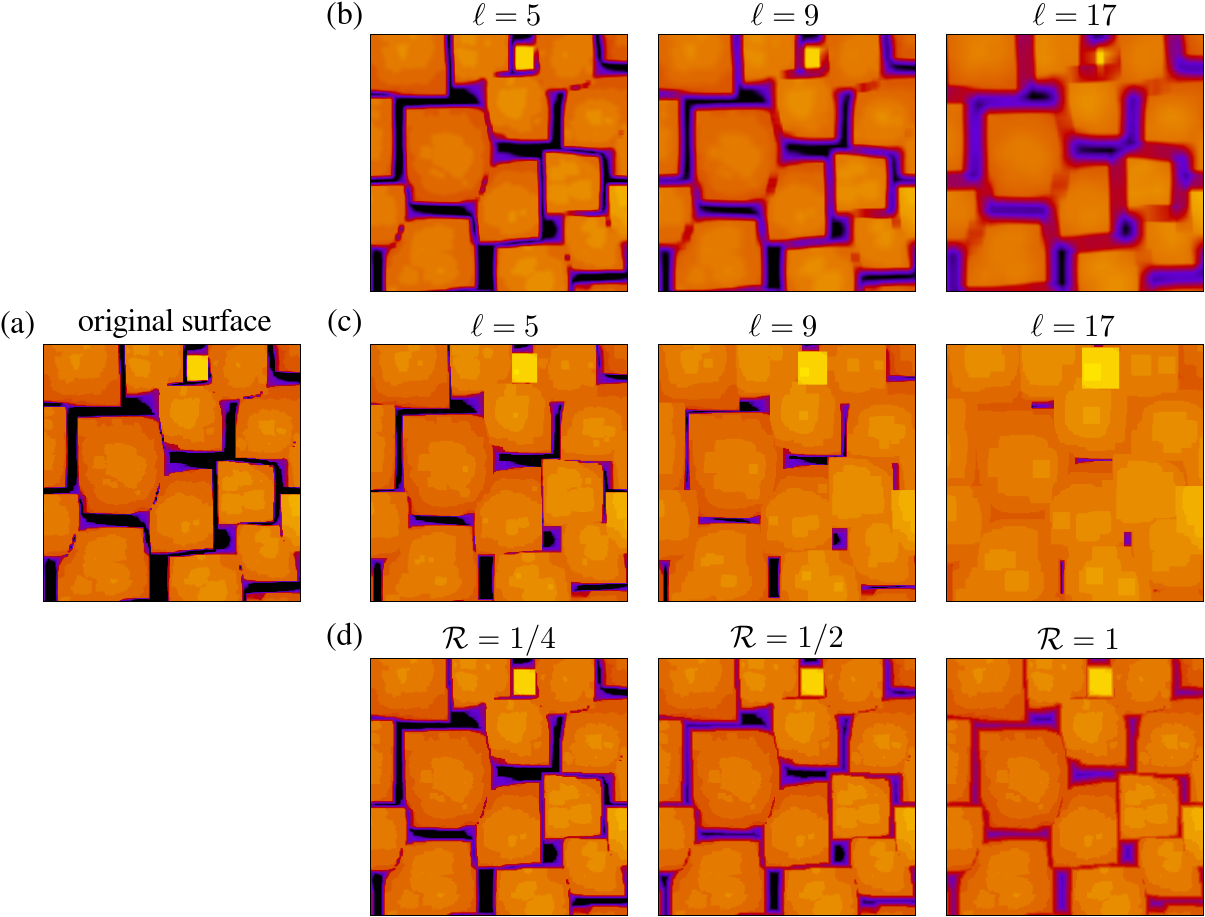}
	\caption{Snapshots of size $256\times256$ taken from: (a) the original surface grown with set D at coverage $\theta = 27.1516$; (b), (c), and (d) the transformed surfaces obtained from (a) using models I, II, and III, respectively.
		The results in (b) and (c) are for square boxes with the indicated sizes.
		Case (d) is for pyramidal tips with base $D=39$ and the indicated aspect ratios.}
	\label{snapshotsD}
\end{figure}

\end{document}